# Forced magnetic reconnection

G. Vekstein[*]

[*]Email address: g.vekstein@manchester.ac.uk

Jodrell Bank Centre for Astrophysics, School of Physics and Astronomy, University of Manchester, Manchester M13 9PL, UK

This is a tutorial-style selective review explaining basic concepts of forced magnetic reconnection. It is based on a celebrated model of forced reconnection suggested by J. B. Taylor. The standard magnetohydrodynamic (MHD) theory of this process has been pioneered by Hahm & Kulsrud (Phys. Fluids **28**, 2412, 1985). Here we also discuss several more recent developments related to this problem. These include energetics of forced reconnection, its Hall-mediated regime, and nonlinear effects with the associated onset of the secondary tearing (plasmoid) instability.

1. Introduction

Magnetic reconnection is a change in connectivity of magnetic field lines taking place in a highly conducting fluid. The very notion of "magnetic reconnection" has been, probably, firstly introduced by J. Dungey (1953). He noticed that magnetic X-type neutral points are structurally unstable: they readily collapse into extended magnetic singularities called current sheets, where magnetic field lines can be broken and re-joined. A detailed theory of this process had been later developed by S. I. Syrovatskii (1971, 1981). Somewhat conceptually close suggestions about a special role of magnetic X-type nulls in the plasma heating and acceleration during solar flares were put forward even earlier by R. Giovanelli (1946) and F. Hoyle (1949). Nowadays, it is firmly established that the process of magnetic reconnection is at the heart of a wide variety of explosive phenomena observed in space and laboratory plasmas (solar coronal activity, magnetospheric substorms, tokamak disruptions, etc., see, e.g. Yamada, Kulsrud & Ji 2010). The simplest way to investigate the process of magnetic reconnection theoretically is to explore models based on the single-fluid resistive magnetohydrodynamics (MHD). Then, temporal evolution of the magnetic field $\vec{B}$ is governed by the following equation:

$$\frac{\partial \vec{B}}{\partial t} = \vec{\nabla} \times (\vec{V} \times \vec{B}) + \eta \nabla^2 \vec{B}, \qquad (1)$$

where $\vec{V}$ is the plasma velocity, and $\eta$ is its magnetic diffusivity. In a magnetically dominated plasma, where the role of reconnection is most important, the thermal pressure $p$ is small compared to the magnetic one (the parameter $\beta \equiv 8\pi p / B^2 \ll 1$), so the characteristic dynamic plasma velocity is the Alfven velocity $V_A = B/\sqrt{4\pi\rho}$. Therefore, in a system with a spatial scale $L$ Eq.(1) defines two different time-scales: the dynamic time-scale $\tau_A = L/V_A$, and the global resistive time $\tau_\eta = L^2/\eta$. The important point is that in almost all cases of interest there is a huge disparity between these two times. Their ratio, called the Lundquist number $S \equiv \tau_\eta / \tau_A = LV_A/\eta$, is very large. For example, in laboratory fusion devices $S \sim 10^6$, while in the solar corona $S \sim 10^{12} - 10^{14}$. Therefore, for processes with a typical duration of $\Delta t \ll \tau_\eta$ one may ignore, as a first step, the resistive term in Eq.(1). This approximation, called the "ideal MHD", made magnetic field frozen-in the plasma flow, which puts severe constraints on the magnetic field evolution. In particular, topology of the magnetic configuration remains preserved (see, e.g. Landau, Lifshitz & Pitaevskii 1984). Although breaking and re-joining of magnetic field lines, i.e. magnetic reconnection, may occur on the time-scale of $t \sim \tau_\eta$, it is of little practical and theoretical interest. Indeed, consider, for example, the case of the coronal active region with $L \sim 10^9 \, cm$ and $V_A \sim 10^3 \, km/s$. This translates into $\tau_\eta \sim 10^{14} s \sim (10^6 \text{years}!)$, while typical solar flare lasts only tens of minutes. Therefore, such a "trivial" magnetic reconnection, when the reconnection time is $\tau_r \sim \tau_\eta$, becomes completely irrelevant.

Thus, in magnetic reconnection research the interest is in the "non-trivial" reconnection process with $\tau_r \ll \tau_\eta$. Clearly, it must involve the resistive term in Eq.(1), which is possible only if at the site of reconnection there is a strong concentration of the electric current, i.e. spatial scale $\Delta L$ of the magnetic field variation there is small: $\Delta L \ll L$. In other words, a current sheet (CS) should be present there. It is worth mentioning that such CS is a necessary pre-requisite not only for the MHD reconnection. At present, there are numerous attempts to develop fast reconnection models by invoking some non-MHD (kinetic) effects such as Hall effect, electron inertia, electron gyroviscosity, etc.(Yamada,

Kulsrud & Ji 2010). However, all of them require some microscale being involved. Thus, in the case of the Hall-mediated reconnection (see Section 4) this is the ion inertial length $d_i = c/\omega_{pi}$, for the electron gyroviscosity-the gyroradius of electrons, for electron inertia – the electron inertial length.

Basically, there are two mechanisms of the CS formation. The first one is associated with some internal MHD instability of a system such as resistive tearing mode (see, e.g. White 1983) or the ideal MHD kink mode (see, e.g. Baty 2000). Therefore, the subsequent magnetic reconnection can be called "spontaneous". However, CS can appear even in an MHD stable magnetic configuration in response to some external perturbation, triggering in this way the so-called "forced" magnetic reconnection. This process is of great interest to astrophysics. Consider, for example, some active region in the solar corona. Its magnetic field is subjected to continuous deformation by photospheric flows that shuffle photospheric footpoints of the field. Since typical time-scale of these flows is about $10^3 s$, which is much longer than the dynamical time $\tau_A \sim 10 s$, the respective perturbation may be considered as a quasistatic (see, e.g. Vekstein 2016). It was conjectured (Parker 1972) that if coronal magnetic field remains frozen-in (which, bearing in mind that $\tau_\eta \sim 10^{14} s$, looks like quite a reasonable assumption), the resulting deformed force-free magnetic equilibrium cannot be smooth: it should contain magnetic singularities in the form of current sheets. Later it was pointed out that essential role in the CS formation is played by a non-trivial structure of the initial magnetic configuration, namely by presence there of the separatrix surfaces (Low & Wolfson 1988; Vekstein, Priest & Amari 1990). The latter does not imply any real limitation for the case of the coronal active region. Its magnetic field acquires quite a complicated structure even when produced by just a few photospheric sources (Gorbachev & Somov 1988). Typically, it contains several topologically distinct domains, boundaries of which intersect along the lines called "separators". External deformations of such a field result in the CS formation along the separators (Longcope & Cowley 1996).

However, CS can form even in a simply-structured magnetic field, if the external perturbation "resonates" with such a field due to some symmetry of the latter. Consider, for example, sheared force-free magnetic field (Bobrova & Syrovatskii 1979):

$$\vec{B}^{(0)} = \{0, B_0 \sin\theta(x), B_0 \cos\theta(x)\}, \qquad (2)$$

where $B_0$ is a constant, so the electric current, which is equal to

$$\vec{j}^{(0)} = \frac{c}{4\pi}(\vec{\nabla} \times \vec{B}^{(0)}) = \frac{cB_0}{4\pi}\frac{d\theta}{dx}\{0, \sin\theta(x), \cos\theta(x)\},$$ is directed along the magnetic field (2). Suppose now that this magnetic field, being immersed into a perfectly conducting plasma, is subjected to some small external perturbation which slightly deforms the initial magnetostatic equilibrium (2): $\vec{B} = \vec{B}^{(0)}(x) + \vec{b}(x,y,z)$, with $b \ll B_0$. In a low-$\beta$ plasma the new equilibrium should be another force-free magnetic field, which requires that

$$\vec{j} \times \vec{B} = (\vec{j}^{(0)} + \vec{j}_1) \times (\vec{B}^{(0)} + \vec{b}) \approx (\vec{j}^{(0)} \times \vec{b}) + (\vec{j}_1 \times \vec{B}^{(0)}) = 0, \qquad (3)$$

with $\vec{j}_1 = c(\vec{\nabla} \times \vec{b})/4\pi$. The magnetic field perturbation, $\vec{b}$, can be represented as $\vec{b}(x,y,z) = \sum \vec{b}(x,k_y,k_z)\exp[i(k_y y + k_z z)]$, and in the adopted linear approximation each Fourier harmonic can be considered separately in Eq.(3). Then, since in the perfectly conducting plasma magnetic field remains frozen-in, by omitting the resistive term in Eq.(1), one gets that

$$\vec{b}(x,\vec{k}) = \vec{\nabla} \times \vec{A}(x,\vec{k}), \ \vec{A}(x,\vec{k}) = \vec{\xi}(x,\vec{k}) \times \vec{B}^{(0)}, \qquad (4)$$

where $\vec{\xi}(x,\vec{k})$ is Fourier harmonic of the plasma displacement $\vec{\xi}(\vec{r}) = \int \vec{V}(\vec{r},t)dt$. It follows now from Eqs.(2) and (4) that

$$A_x = \xi_\perp B_0, A_y = -\xi_x B_0 \cos\theta, A_z = \xi_x B_0 \sin\theta, \qquad (5)$$

with $\xi_\perp = \xi_y \cos\theta - \xi_z \sin\theta$ being plasma displacement component perpendicular to the initial magnetic field in the $(y-z)$ plane. Thus, pretty straightforward but quite cumbersome derivations involving Eqs.(3-5) yield that the sought after deformed equilibrium is provided by the displacements $\xi_x(x,\vec{k})$ and $\xi_\perp(x,\vec{k})$ that are solutions of the following equations:

$$\frac{d^2\xi_x}{d^2 x} + \frac{2k_\perp}{k_{II}}\frac{d\theta}{dx}\frac{d\xi_x}{dx} - k^2\xi_x = 0, \ \xi_\perp = i\frac{k_\perp}{k^2}\frac{d\xi_x}{dx}. \qquad (6)$$

Here $k_{II}(x) = k_y \sin\theta + k_z \cos\theta$ and $k_\perp(x) = k_y \cos\theta - k_z \sin\theta$ are components of $\vec{k}$ directed, respectively, along and across the initial magnetic field at a given location $x$. As seen from (6), the equation for $\xi_x$ acquires a singularity at some $x = x_0$, where $k_{II}(x_0) = 0$, i.e. $\vec{k} \cdot \vec{B}^{(0)}(x_0) = 0$, which means that at this location the perturbation does not vary along the initial magnetic field line. Thus, the plane $x = x_0$ is the resonant surface where the CS could be expected to form. At this location there is a reversal of the magnetic field lines projected into the $(\vec{x}, \vec{k})$ plane. The electric current associated with the CS is directed perpendicular to this plane, and, hence, along the initial magnetic field at $x = x_0$ (as it requires for the force-free equilibrium). This process is considered in detail in Section 2.

Note, however, that there is nothing special with such a plane itself: any plane $x = const$ can be made a resonant surface with the appropriate choice of the perturbation vector $\vec{k} = (k_y, k_z)$ (presuming, of course, that the electric current there is not equal to zero, i.e. $d\theta/dx \neq 0$). Such resonant surfaces are also relevant to spontaneous magnetic reconnection such as the tearing instability (Furth, Killeen & Rosenbluth 1963): CS is formed there in the course of the instability development.

A similar effect takes place also in toroidal magnetic configurations. Thus, assume a tokamak-type toroidal camera (see, e. g. Boozer 2004) with a major radius $R$ and a minor radius $a \ll R$. It contains almost uniform toroidal magnetic field $B_\theta$ produced by external coils, and poloidal field $B_\phi(r)$ generated by the plasma toroidal electric current. Thus, in the ideal situation the tokamak magnetic configuration would be a superposition of nested toroidal magnetic surfaces. However, in the real world some additional unwanted albeit small "error" fields are inevitable (for example, due to discreteness of toroidal coils). They can be represented as a superposition of toroidal/poloidal perturbation harmonics as $\delta \vec{B} = \sum \vec{b}_{nm}(r) \exp[i(n\theta - m\phi)]$. Consider now a single harmonic with some toroidal/poloidal numbers $(n, m)$. Since along a magnetic field line of the unperturbed field $\frac{rd\phi}{B_\phi} = \frac{Rd\theta}{B_\theta}$, the respective perturbation does not vary along this line of force, i.e. $nd\theta - md\phi = 0$, if $q(r) \equiv \frac{rB_\theta}{RB_\phi} = \frac{m}{n}$ [The parameter $q(r)$ is known as a "safety factor" because of its relevance to MHD stability of the system (see, e. g. Boozer 2004)]. Thus, resonant magnetic surfaces are those for which $q(r)$ is a rational number. Therefore, error fields can drive there forced reconnection of magnetic field lines, which severely degrade global energy confinement in the tokamak device (see, e.g., Fitzpatrick 2010).

The rest of the paper is organised as follows. In Section 2 we introduce a force-free modification of the Taylor's model, and discuss its energetics. The latter issue is related to the MHD stability of the system, which is considered in Section 3. Section 4 is devoted to the standard single-fluid MHD linear theory of forced reconnection, while its Hall-mediated counterpart is presented in Section 5. Finally, nonlinear effects, and the associated onset of plasmoid instability are discussed in Section 6.

## 2. The Taylor's problem and energetics of forced magnetic reconnection

Consider now a linear force-free magnetic field of type (2), for which the shear parameter is uniform: $d\theta/dx = const = \alpha$, hence

$$\vec{B}^{(0)} = \{0, B_0 \sin \alpha x, B_0 \cos \alpha x\}. \tag{7}$$

This field is embedded in a zero-$\beta$ plasma and bounded by two perfectly conducting surfaces located at $x = x_b^{(\pm)} = \pm l$. Then this system is subjected to external perturbation which comprises a small bending of the boundary surfaces as

$$x_b^{(\pm)} = \pm(l + a\cos ky), a/l << 1 \tag{8}$$

Clearly, in the new force-free equilibrium the magnetic field will be different from the initial field (7) since magnetic field lines should remain tangential to the perfectly conducting boundary surfaces. Since the system remains invariant in the z-direction, in order to derive the deformed magnetic equilibrium it is helpful to introduce the poloidal magnetic flux function $\Psi(x, y)$ (see, e.g. Vekstein 2016) as

$$\vec{B}(x, y) = (\vec{\nabla}\Psi(x, y) \times \vec{z}) + B_z(x, y)\vec{z} \tag{9}$$

Then, in the linear approximation with respect to the small parameter $a/l$, one can represent it in the form

$$\Psi(x, y) = \Psi_0(x) + \psi(x)\cos ky, \tag{10}$$

where $\Psi_0(x) = \dfrac{B_0}{\alpha}\cos\alpha x$ corresponds to the initial field (7), while the second term on the r.h.s. of (10) is due to the perturbation (8). Hence, $\psi(x)$ must be an even function of $x$, the boundary condition for which comes from the requirement that the deformed boundaries remain flux-surfaces, i.e. $\Psi(x = x_b, y) = const$. Thus, in the linear approximation one gets:

$$\Psi_0(l) + \left.\dfrac{d\Psi_0}{dx}\right|_{x=l} a\cos ky + \psi(l)\cos ky = const \Rightarrow \psi(l) = B_0 a \sin\alpha l \tag{11}$$

The flux-function (10) of the deformed force-free magnetic equilibrium should satisfy the Grad-Shafranov equation (see, e.g. Vekstein 2016)

$$B_z(x, y) \equiv B_z(\Psi), \quad \nabla^2\Psi + B_z\dfrac{dB_z}{d\Psi} = 0 \tag{12}$$

Then, since the initial field (7) is a linear force-free field for which $B_z(\Psi) = \alpha\Psi$, the same remains true for the perturbed equilibrium too (Vekstein & Jain 1998). Hence, Eq. (12) reduces to $\nabla^2\Psi + \alpha^2\Psi = 0$, which yields the following equation for the function $\psi(x)$ in (10):

$$\dfrac{d^2\psi}{d^2x} + (\alpha^2 - k^2)\psi = 0 \tag{13}$$

As shown below, a long-wave perturbation, with $k < \alpha$, is of primary interest, so in what follows it is assumed that $\kappa^2 \equiv (\alpha^2 - k^2) > 0$. In this case, a regular solution of Eq.(13) with the boundary condition (11) reads

$$\psi(x) = \psi^{(r)}(x) = B_0 a \dfrac{\sin\alpha l}{\cos\kappa l}\cos\kappa x \tag{14}$$

It turns out, however, that such a seemingly simple solution results in the equilibrium whose magnetic field lines topology is different from that of the initial equilibrium (2). As shown in Fig.1a, the so-called "magnetic islands" are formed in a vicinity of the plane $x = 0$. Indeed, by using expressions (10) and (14), at the vicinity of $x = 0$ the poloidal flux function can be approximated as

$$\Psi(x, y) \approx \dfrac{B_0}{\alpha}(1 - \alpha^2 x^2/2) + \psi^{(r)}(0)\cos ky \tag{15}$$

(note that $\psi^{(r)}(0) = B_0 a \frac{\sin \alpha l}{\cos \kappa d} \neq 0$!). It follows then from (15) that the point $(0,0)$ with $\Psi(0,0) = \frac{B_0}{\alpha} + \psi^{(r)}(0)$ corresponds to the O-point inside the magnetic island, while the X-point on the separatrix has the coordinates $(0, \pi/k)$ and the flux function value of $\Psi(0, \pi/k) = \Psi_s = \frac{B_0}{\alpha} - \psi^{(r)}(0)$. Thus, reconnected magnetic flux inside each island, $(\Delta \psi)_r$ is equal to

$$(\Delta \psi)_r = 2\psi^{(r)}(0) = 2B_0 a \frac{\sin \alpha l}{\cos \kappa d} \tag{16}$$

The island's width, $w$, is equal to $2x_s$, where $x_s$ is the x-coordinate of the separatrix point with $y_s = 0$ (see Fig.1a). It follows then from (15) that

$$\Psi(x_s, 0) = \frac{B_0}{\alpha}(1 - \alpha^2 x_s^2 / 2) + \psi^{(r)}(0) = \Psi_s = \frac{B_0}{\alpha} - \psi^{(r)}(0), \text{ which yields}$$

$$w_r = 4[\psi^{(r)}(0) / \alpha B_0]^{1/2} \tag{17}$$

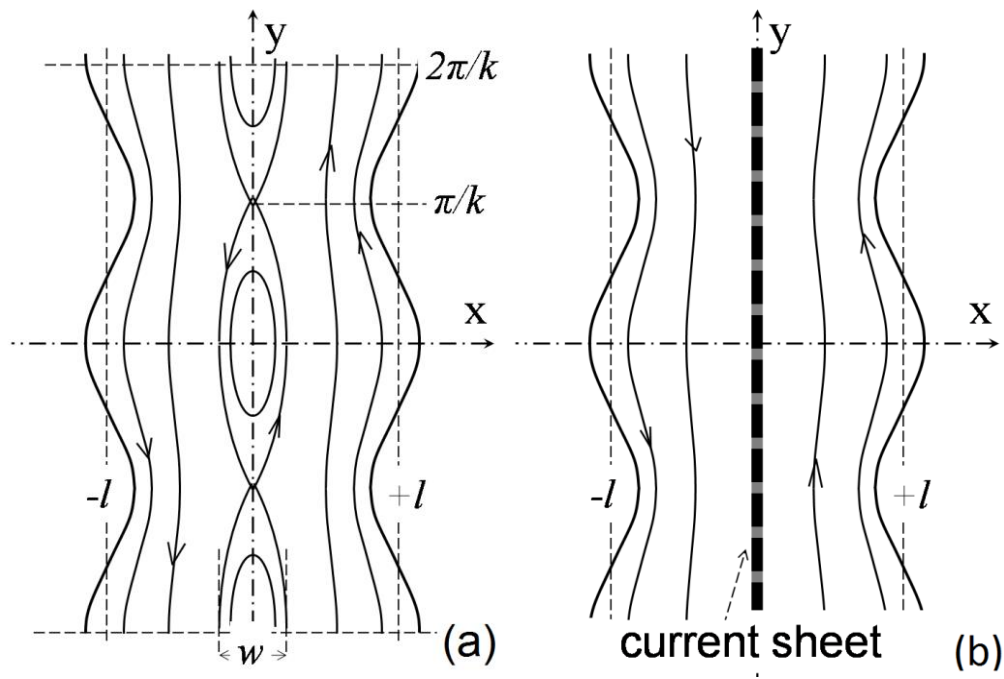

Figure 1. New MHD equilibria: (a) – reconnected equilibrium with magnetic islands; (b) – ideal MHD equilibrium with the current sheet.

Clearly, such equilibrium cannot form in a perfectly conducting plasma, where the topology of magnetic field lines is preserved due to the frozen-in constraint. Therefore, another, ideal MHD equilibrium, that does not contain magnetic islands, should exist. Such a requirement implies that the respective solution of Eq.(13), $\psi^{(i)}(x)$, must satisfy an additional condition $\psi^{(i)}(0) = 0$ and, therefore, it cannot be a regular one (one too many boundary conditions). Thus, one gets

$$\psi^{(i)}(x) = B_0 a \frac{\sin \alpha l}{\sin \kappa d} |\sin \kappa x| \qquad (18)$$

This is a singular solution that acquires the discontinuity at $x = 0$: a current sheet across which the magnetic field component $B_y = -\partial \Psi / \partial x$ has a finite jump

$$\{B_y\} \equiv B_y\big|_{0+} - B_y\big|_{0-} = -2B_0 a\kappa \frac{\sin \alpha l}{\sin \kappa d} \qquad (19)$$

In this particular case the wave vector of the perturbation (8) is equal to ($k_y = \mp k, k_z = 0$), so the location of the CS is indeed the resonant surface where $\vec{k} \cdot \vec{B}^{(0)} = 0$ (see Section 1). This ideal MHD equilibrium is depicted in Fig.1b. Note that the respective surface electric current, $i_z = c\{B_y\}/4\pi$, is directed along the initial magnetic field (7), so the overall magnetic equilibrium is not violated.

A question of how these two solutions manifest themselves when the plasma resistivity is very small but finite was raised by J. B. Taylor (hence, Taylor's problem). It has been considered in a seminal paper by Hahm & Kulsrud (1985), who demonstrated that forced magnetic reconnection is a process of the transition from the ideal MHD equilibrium to the reconnected one, which takes place at several stages. This issue will be discussed in detail in Section 4. The Taylor problem, being by itself of considerable interest in the plasma physics theory, has also many important implications for laboratory and astrophysical plasmas (see, e.g., Comisso et al 2015, and a list of references therein). In the context of astrophysics the most important aspect of forced magnetic reconnection is its energetics. For example, in the solar corona forced reconnection can be triggered by the motion of the photospheric footpoints of the coronal magnetic field. The result is a fast release of excess magnetic energy stored in the corona, which causes such observable explosive coronal events as solar flares and coronal mass ejections.

Thus, consider now the energetics of forced reconnection for the particular case of the Taylor model. In order to do so, one should compare total magnetic energies of the following three equilibria: the initial field (7), the ideal MHD state with the respective flux function perturbation (18), and the reconnected equilibrium described by Eq.(14). As far as the magnetic energy of the initial field is concerned, the result is trivial: calculated per unit area in the $(y-z)$ plane, it is equal to $W_M^{(0)} = B_0^2 l / 4\pi$. In the case of the perturbed equilibria the respective energy can be also written as $W_M = (8\pi)^{-1} \int_{-l}^{+l} \langle B^2 \rangle dx$, where symbol $\langle \ \rangle$ means averaging over the variation along the y-coordinate. However, deriving the energy in this way is not that simple. The point is that a change of the magnetic energy for the perturbed equilibria is of the order of $(a/l)^2$. Therefore, such a direct calculation would require knowledge of the second order corrections to the deformed magnetic field. This quite a cumbersome task can be bypassed with help of the consideration that follows a spirit of the so-called "energy principle" (Bernstein 1984 ). Thus, assume that the boundary deformation (8) is caused by some external force in a gradual way, so that the perturbation amplitude, $\delta(t)$, is slowly increasing from zero to its terminal value equal to $a$. In the course of this quasistatic deformation the internal magnetic field should remain force-free at any instant of time, with the external force being balanced by the internal magnetic pressure at the boundary surface. Therefore, the sought after change in the internal magnetic energy is equal to the work of the external force providing the boundary deformation:

$$\Delta W_M = -2 \int_0^a \left\langle \frac{B^2(x = x_b^{(+)})}{8\pi} \cos ky \right\rangle d\delta \qquad (20)$$

(a factor of 2 in (20) accounts for the second boundary located at $x = x_b^{(-)}$).

As seen from (20), in performing this derivation it is sufficient to use the magnetic flux perturbations $\psi^{(i,r)}$ already obtained in the linear approximation [(see Eqs. (14) and (18)], where the amplitude $a$ is replaced with $\delta(t)$.

Thus, the procedure is as follows. The deformed magnetic field is $\vec{B} = \vec{B}^{(0)} + \vec{b}$, where the field perturbation $\vec{b}$ is related to the flux function perturbation $\psi_1 \equiv \psi(x)\cos ky$ in Eq.(10) as

$$b_x = \frac{\partial \psi_1}{\partial y} = -k\psi(x)\sin ky, \quad b_y = -\frac{\partial \psi_1}{\partial x} = -\psi'(x)\cos ky, \quad b_z = \alpha\psi_1 = \alpha\psi(x)\cos ky$$

Then, the magnetic pressure at the deformed boundary, calculated in the required linear approximation, takes the form

$$\frac{B^2(x_b^{(+)})}{8\pi} = \frac{B_0^2}{8\pi} + \frac{B_0}{4\pi}[\alpha\psi(l)\cos\alpha l - \psi'(l)\sin\alpha l]\cos ky \qquad (21)$$

Obviously, only the last term in (21) makes a non-zero contribution in (20). Then, after using expressions (14) and (18) for $\psi^{(i,r)}$ [with $a$ replaced there by $\delta(t)$], Eqs. (20-21) yield the following magnetic energies for the two deformed equilibria:

$$W_M^{(i)} = W_M^{(0)} + \Delta W_M^{(i)} = W_m^{(0)} + \frac{B_0^2 a^2 \sin^2(\alpha l)}{8\pi d}[(\kappa d)\cot(\kappa d) - (\alpha l)\cot(\alpha l)] > W_M^{(0)} \quad (22a)$$

$$W_M^{(r)} = W_M^{(0)} + \Delta W_M^{(r)} = W_m^{(0)} - \frac{B_0^2 a^2 \sin^2(\alpha l)}{8\pi d}[(\kappa d)\tan(\kappa d) + (\alpha l)\cot(\alpha l)] < W_M^{(0)} \quad (22b)$$

Therefore, the net energy released in the process of forced magnetic reconnection is equal to

$$\Delta W_R = W_M^{(i)} - W_M^{(r)} = \frac{B_0^2 a^2 \sin^2(\alpha l)}{8\pi d}(\kappa d)[\cot(\kappa d) + \tan(\kappa d)] > 0 \qquad (23)$$

As seen from (22a), such external perturbation moves total magnetic energy of the system slightly up ($\Delta W_M^{(i)} > 0$). However, according to (23), the energy released by the subsequent magnetic reconnection exceeds the extra magnetic energy supplied externally because the terminal reconnected state has lower magnetic energy than the initial field ($\Delta W_M^{(r)} < 0$). Therefore, forced reconnection can be viewed as a mechanism of the internal magnetic relaxation (Vekstein & Jain 1998). Moreover, a difference between the externally supplied and released energies can be very large, so the role of the external perturbation is just to trigger the reconnection process, while the released energy is mainly tapped from the excess magnetic energy stored in the initial magnetic configuration. This is clearly the case when the parameter $\kappa d$ is close to $\pi/2$ (note that the second term on the r.h.s. of Eq.(23) diverges when $\kappa d \to \pi/2$). The reason is that the field (7) becomes tearing unstable if $\kappa d = (\alpha^2 - k^2)^{1/2} l > \pi/2$ (see Section 3). Thus, modes with $kl \to 0$ are most unstable, which yields the overall instability threshold for the shear parameter of this field: $\alpha > \alpha_{cr} = \pi/2l$. Clearly, the very issue of forced magnetic reconnection makes sense only when the initial state is MHD stable. Therefore, in what follows it is assumed that $\alpha < \pi/2l$.

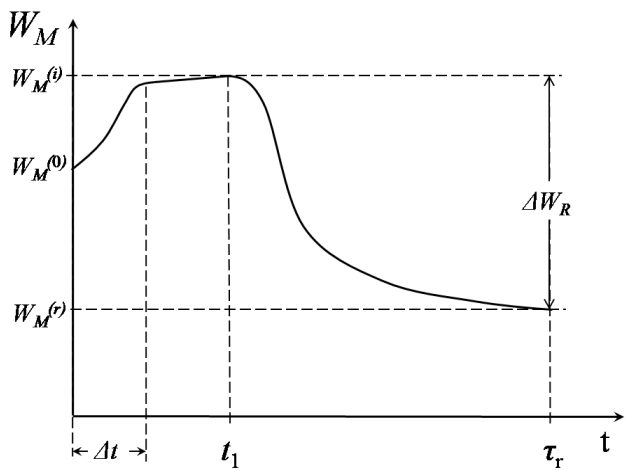

Figure 2. Reconnective magnetic relaxation of the system.

The above mentioned enhancement of the released energy in the marginally stable field is due to a very large size of magnetic islands in this limit [see Eqs. (16) and (17)], which makes the linear approximation explored so far not applicable even for a small perturbation amplitude $a$. Taking into account a finite size of magnetic islands saturates the released energy (Vekstein 2004), which, nevertheless, still by far exceeds the extra energy provided by the external source (Browning et al 2001). An overall scenario of the magnetic energy temporal evolution in the process of forced reconnection is sketched in Fig.2. A system undergoes external perturbation lasting several Alfven times: $\Delta t \sim \tau_A = l/V_A$. At this stage a small plasma resistivity still plays no role, so the ideal MHD equilibrium with the current sheet and the magnetic energy equal to $W_M^{(i)}$ is formed. Then, on a much longer time scale $\tau_r \gg \tau_A$ (see Section 4), resistive effects intervene, eventually destroying the current sheet and bringing about transition to the reconnected equilibrium with a different magnetic topology and lower magnetic energy, $W_M^{(r)}$.

3. **MHD stability of a planar force-free magnetic field**.

Consider now MHD stability of the planar force-free magnetic field (2), starting with the case of the ideal MHD. Then, by writing $\vec{B} = \vec{B}^{(0)} + \vec{b}$, the magnetic field perturbation $\vec{b}$ in the linear approximation is related to the plasma displacement vector $\vec{\xi}$ according to Eq.(4). Therefore, the evolution equation for $\vec{\xi}$ (the equation of motion) takes the form

$$\rho \frac{\partial^2 \vec{\xi}}{\partial^2 t} = \frac{1}{c} \vec{j} \times \vec{B} = \frac{1}{c}[(\vec{j}^{(0)} + \vec{j}_1) \times (\vec{B}^{(0)} + \vec{b})] \approx \frac{1}{c}[(\vec{j}^{(0)} \times \vec{b}) + (\vec{j}_1 \times \vec{B}^{(0)})] \equiv \vec{F}(\vec{\xi}), \quad (24)$$

where the force $\vec{F}(\vec{\xi})$ is a linear functional of the displacement $\vec{\xi}(\vec{r})$. Here

$$\vec{j}^{(0)} = \frac{c}{4\pi}(\vec{\nabla} \times \vec{B}^{(0)}) = \frac{cB_0}{4\pi} \frac{d\theta}{dx} \vec{B}^{(0)}, \quad \vec{j}_1 = \frac{c}{4\pi}(\vec{\nabla} \times \vec{b}). \quad (25)$$

Thus, following the "energy principle" (Bernstein 1984), one can introduce the "potential energy" as $U = -\frac{1}{2}\int \vec{F}(\vec{\xi}) \cdot \vec{\xi} dV$, which is a quadratic of $\vec{\xi}$. Then, the system is MHD stable if the potential energy $U \geq 0$ for any permissible displacement field $\vec{\xi}(\vec{r})$. It follows from (24-25) that

$$U = \frac{1}{8\pi}\int dV \vec{\xi} \cdot \{[(\vec{\nabla} \times \vec{b}) \times \vec{B}^{(0)}] + [(\vec{\nabla} \times \vec{B}^{(0)}) \times \vec{b}]\},$$

which, by integrating the first term by parts and using the boundary condition $\vec{\xi} \cdot d\vec{S} = 0$, reduces to

$$U = \frac{1}{8\pi}\int dV\{[\vec{\nabla} \times (\vec{\xi} \times \vec{B}^{(0)})]^2 - [\vec{\xi} \times (\vec{\nabla} \times \vec{B}^{(0)})] \cdot [\vec{\nabla} \times (\vec{\xi} \times \vec{B}^{(0)})]\} \quad (26)$$

Since the first term on the r.h.s. of (26) is always non-negative, the MHD instability, if any, should be associated with the electric current present in the initial magnetic configuration, $(\vec{\nabla} \times \vec{B}^{(0)} \neq 0)$. This is a demonstration that such a current is a source of excess magnetic energy that can make the system unstable (see, e.g. Vekstein 2016).

In the particular case under consideration the vector potential perturbation $\vec{A}$ is given by Eq.(5). Then, without any loss of generality, one can represent components of the displacement $\vec{\xi}(\vec{r})$ as

$$\xi_x = \xi_1(x)\cos ky, \quad \xi_\perp = \xi_2(x)\cos ky + \xi_3(x)\sin ky,$$

after which the magnetic field perturbation $\vec{b} = \vec{\nabla} \times \vec{A}$ takes the form

$$b_x = -k\xi_1 B_0 \sin\theta \sin ky, \quad b_y = -B_0(\xi_1' \sin\theta + \xi_1 \theta' \cos\theta)\cos ky,$$
$$b_z = B_0(\xi_1 \theta' \sin\theta - \xi_1' \cos\theta)\cos ky + kB_0(\xi_2 \sin ky - \xi_3 \cos ky) \qquad (27)$$

Then, the potential energy (26), which is now derived per unit area in the (y-z) plane, can be written as $U = \dfrac{1}{8\pi}\int_{-l}^{l} dx[\langle b^2 \rangle - \theta'\langle \vec{A}\cdot\vec{b}\rangle]$. By making use of expressions (5) and (27), a straightforward calculation results in

$$U = \frac{B_0^2}{16\pi}\int_{-l}^{l} dx[k^2\xi_1^2(\sin\theta)^2 + k^2(\xi_2^2 + \xi_3^2) + (\xi_1')^2 - 2k\xi_1'\xi_3\cos\theta]$$

Its minimization with respect to $\xi_2$ and $\xi_3$ yields $\xi_2 = 0, \xi_3 = -\xi_1'(\cos\theta)/k$, so finally one gets

$$U = \frac{B_0^2}{8\pi}\int_{-l}^{l} dx(\sin\theta)^2[(\xi_1')^2 + k^2\xi_1^2] \geq 0 \qquad (28)$$

Therefore, this magnetic configuration is stable in the framework of the ideal MHD for any shear function $\theta(x)$.

However, it may become unstable for a wider class of perturbations that are not allowed in the ideal MHD with its frozen-in magnetic field restriction. This can be demonstrated (Goedbloed & Dagazian 1971) by considering the potential energy (28) expressed in terms of the flux function perturbation $\psi_1 = \psi(x)\cos ky$ introduced in Eq.(10). Since this flux function is equivalent to the vector potential component $A_z$, it follows from (5) that the displacement $\xi_1(x) = \psi(x)/B_0\sin\theta(x)$, so a simple integration by parts in (28) results in

$$U = \frac{1}{8\pi}\int_{-l}^{l} dx[(\psi')^2 - (\alpha^2 - k^2)\psi^2] \qquad (29)$$

(for simplicity, here and in what follows it is assumed that the initial planar force-free field has a uniform shear, i. e. $\theta(x) = \alpha x$). Clearly, if such instability does exist, it requires a deviation from the ideal MHD framework, i.e. some finite plasma resistivity, $\eta$, should be present. Therefore, such an approach, which is based on the energy principle, is valid only when the resistive effects are weak enough. In quantitative terms it means that the respective Lundquist number, $S \equiv lV_A/\eta$, is large: $S \gg 1$. It is worth reminding here (see Section 1) that this is the case for almost all applications of interest. Then, development of the resulting instability is quite slow: the instability growth rate $\gamma \ll \tau_A^{-1}$, which means that the process is quasistatic. Therefore, the system remains close to the magnetostatic equilibrium, hence, the flux function perturbation $\psi$ that defines the potential energy in Eq.(29), should be a solution of Eq.(13). Note that unlike the problem of forced magnetic reconnection discussed in Section 2, in this case the boundary surfaces $x = \pm l$ are not deformed, so the flux function perturbation should vanish there: $\psi(x = \pm l) = 0$. A regular solution of (13) with such boundary conditions, which is $\psi(x) \equiv 0$, cannot lead to instability that requires $U < 0$.

However, a negative $U$ can be achieved if the function $\psi(x)$ has a discontinuous first derivative. Since the latter is proportional to $b_y$, the y-component of the magnetic field perturbation, such a singularity implies a current sheet with a surface electric current flowing along the z-axis. Therefore, such a current sheet should be located where this current is directed along the initial magnetic field $\vec{B}^{(0)}(x)$, i. e. at $x = 0$ (otherwise, an infinite magnetic force would be exerted on the plasma). Thus, bearing this in mind, one can transform expression (29) for the potential energy with help of the integration by parts in the intervals $(-l, 0-\varepsilon)$ and $(0+\varepsilon, l)$ with $\varepsilon \to 0$ as

$$U = \frac{1}{8\pi} \left\{ \int_{-l}^{0-\varepsilon} dx \left[ (\psi')^2 + (k^2 - \alpha^2)\psi^2 \right] + \int_{0+\varepsilon}^{l} dx \left[ (\psi')^2 + (k^2 - \alpha^2)\psi^2 \right] \right\} =$$

$$= \frac{1}{8\pi} \left\{ -\int_{-l}^{0-\varepsilon} dx \left[ \psi'' + (\alpha^2 - k^2)\psi \right] \psi + (\psi\psi')\big|_{-l}^{0-\varepsilon} + (\psi\psi')\big|_{0+\varepsilon}^{l} - \int_{0+\varepsilon}^{l} dx \left[ \psi'' + (\alpha^2 - k^2)\psi \right] \psi \right\}$$

If the flux function satisfies the equilibrium equation (13) with the boundary conditions $\psi(x = \pm l) = 0$, the above expression yields

$$U = -\frac{1}{8\pi}\psi(0)\left[\psi'\big|_{0-\varepsilon}^{0+\varepsilon}\right] = -\frac{1}{8\pi}\Delta'[\psi(0)]^2, \quad \text{where}$$

$$\Delta' = \frac{\left[\psi'\big|_{0-\varepsilon}^{0+\varepsilon}\right]}{\psi(0)} \qquad (30)$$

The following note is due here. Of course, the required finite plasma resistivity smooths this discontinuity, which results in a finite current sheet width, $\Delta x$. Nevertheless, under large $S$ the CS width is small, $\Delta x \ll l$ (see below), therefore, it has little effect on the derived potential energy (30).

Thus, the conclusion is that, if $\Delta' > 0$, such a perturbation reduces magnetic energy of the system and, hence, can lead to its MHD instability. The essential point here is that the respective energy reduction requires $\psi(0) \neq 0$. However, as it has been demonstrated in Section 2, this condition is not compatible with the ideal MHD, because it results in the change of the magnetic field topology due to formation of magnetic islands. Therefore, this instability, which is called the **tearing mode** (Furth et al 1963), cannot develop without a finite plasma resistivity. However, the latter has no effect on the instability threshold, which is determined entirely by the sign of the parameter $\Delta'$. Note also that the tearing mode theory (see, e.g. White 1983) yields the growth rate $\gamma \sim \tau_A^{-1} S^{-3/5}$ and the current sheet width $\Delta x \sim l S^{-2/5}$, which justify validity of the energy principle consideration in this case.

Now one can apply this recipe to the initial magnetic field (7). The first step is derivation of a non-trivial solution of Eq.(13), which satisfies the boundary conditions $\psi(x = \pm l) = 0$ but has a discontinuous first derivative at $x = 0$. Assuming first that $\kappa^2 = \alpha^2 - k^2 > 0$, one gets

$$\psi(x) = \psi(0)\sin[\kappa(x+l)]/\sin\kappa l, -l \leq x \leq 0; \quad \psi(x) = \psi(0)\sin[\kappa(x-l)]/\sin\kappa l, 0 \leq x \leq l$$

This yields $\Delta' = -2\kappa \cot \kappa l$, so the field (7) becomes tearing unstable, i.e. $\Delta' > 0$, when $\kappa l = (\alpha^2 - k^2)^{1/2} l > \pi/2$. Therefore, perturbations with $kl \to 0$ are most unstable, which leads to the following overall instability threshold: $\alpha > \alpha_{cr} = \pi/2l$. On the other hand, in the case of $\alpha^2 - k^2 < 0$, a simple replacement of $\kappa$ with $i\kappa$ in the above-given derivation yields $\Delta' = -2cth(\kappa l) < 0$, which demonstrates tearing stability of all short wave-length ($k > \alpha$) perturbations.

## 4. **Dynamics of forced reconnection: the single-fluid MHD**.

Consider now how the process of forced magnetic reconnection proceeds with time (Hahm & Kulsrud 1985) after the initial magnetic field (7) underwent the boundary surface deformation (8). In the case of a large Lundquist number, $S = (lV_A/\eta) \gg 1$, a key role is played here by the current sheet (CS) associated with the ideal MHD equilibrium (18). Within this CS plasma dynamics is governed by the induction equation (1) and the equation of motion

$$\rho \frac{d\vec{V}}{dt} = \frac{1}{c}(\vec{j} \times \vec{B}) = -\frac{1}{4\pi} \vec{B} \times (\vec{\nabla} \times \vec{B}) \qquad (31)$$

(it is assumed that plasma thermal pressure is negligibly small). Then, by representing the magnetic field and plasma velocity (the latter is almost incompressible because a strong guide field $B_z^{(0)} \approx B_0$ is present in the CS) as $\vec{B}(x,y,t) = (\vec{\nabla}\Psi \times \vec{z}) + B_z\vec{z}$, $\vec{V}(x,y,t) = \vec{\nabla}\Phi \times \vec{z}$, where $\Phi$ is a stream-function of the flow, these equations take the form

$$\frac{\partial \Psi}{\partial t} = (\vec{\nabla}\Psi \times \vec{\nabla}\Phi) \cdot \vec{z} + \eta \nabla^2 \Psi, \qquad (32a)$$

$$\frac{\partial B_z}{\partial t} = (\vec{\nabla}B_z \times \vec{\nabla}\Phi) \cdot \vec{z} + \eta \nabla^2 B_z, \qquad (32b)$$

$$\rho \frac{d}{dt}(\nabla^2 \Phi) = -\frac{1}{4\pi}[\vec{\nabla}\Psi \times \vec{\nabla}(\nabla^2\Psi)] \cdot \vec{z} \qquad (32c)$$

In the linear approximation, when $B_z = B_z^{(0)}(x) + B_z^{(1)}(x,y,t)$ and $\Psi = \Psi_0(x) + \Psi_1(x,y,t)$ with $B_z^{(0)}$ and $\Psi_0$ given by equations (7) and (10), one gets from (32):

$$\frac{\partial \Psi_1}{\partial t} = -\frac{d\Psi_0}{dx} \cdot \frac{\partial \Phi}{\partial y} + \eta \nabla^2 \Psi_1, \qquad (33a)$$

$$\frac{\partial B_z^{(1)}}{\partial t} = -\frac{dB_z^{(0)}}{dx} \cdot \frac{\partial \Phi}{\partial y} + \eta \nabla^2 B_z^{(1)}, \qquad (33b)$$

$$\frac{\partial}{\partial t}(\nabla^2 \Phi) = -\frac{1}{4\pi\rho}\left[\frac{d\Psi_0}{dx} \cdot \frac{\partial}{\partial y}(\nabla^2 \Psi_1) - \frac{d^3\Psi_0}{d^3x} \cdot \frac{\partial \Psi_1}{\partial y}\right] \qquad (33c)$$

Since the CS thickness, $\Delta x$, is small compared with the system size $l$ and the perturbation wave-length $\lambda = 2\pi/k$, in what follows one can simplify $\frac{d\Psi_0}{dx}, \frac{dB_z^{(0)}}{dx} \propto \sin\alpha x \approx \alpha x$, and assume that $\nabla^2(\Psi_1, B_z^{(1)}, \Phi) \approx \partial^2(\Psi_1, B_z^{(1)}, \Phi)/\partial^2 x$.

Furthermore, it is useful to introduce non-dimensional variables and functions by re-scaling them as follows: length- with $l$, time – with the Alfven time-scale $\tau_A = l/V_A = l\sqrt{4\pi\rho}/B_0$, the flux-function perturbation $\Psi_1$ – with $lB_0(a/l) = aB_0$, $B_z^{(1)}$ - with $B_0(a/l)$, and the stream-function $\Phi$ - with $lV_A(a/l) = aV_A$. These yield the following transformation of Eqs. (33):

$$\frac{\partial \Psi_1}{\partial t} \approx \varepsilon x \frac{\partial \Phi}{\partial y} + \frac{1}{S}\left(\frac{\partial^2 \Psi_1}{\partial^2 x}\right), \qquad (34a)$$

$$\frac{\partial B_z^{(1)}}{\partial t} \approx \varepsilon^2 x \frac{\partial \Phi}{\partial y} + \frac{1}{S}\frac{\partial^2 B_z^{(1)}}{\partial^2 x}, \qquad (34b)$$

$$\frac{\partial}{\partial t}\left(\frac{\partial^2 \Phi}{\partial^2 x}\right) \approx \varepsilon x \frac{\partial}{\partial y}\left(\frac{\partial^2 \Psi_1}{\partial^2 x}\right), \qquad (34c)$$

where $\varepsilon = \alpha l$. According to Chapter 3, this parameter determines tearing stability of the initial magnetic field (7). Therefore, hereafter it is assumed, that, by the very meaning of forced reconnection, the field (7) is MHD stable, i.e. $\varepsilon < \varepsilon_{cr} = \pi/2$, and also it is not too close to the instability threshold, so that expression (14) for the reconnected state, obtained in the linear approximation, still holds. Therefore, in all estimates given below it is assumed that the parameter $\varepsilon \sim 1$.

The variation of the flux-function $\Psi_1$ along the y-axis and its symmetry is imposed by the boundary deformation (8): $\Psi_1(x,y,t) = \psi(x,t)\cos ky$, with $\psi(x,t)$ being an even function of $x$. Then, according to (34c), $\Phi(x,y,t) = \phi(x,t)\sin ky$, where $\phi(x)$ is an odd function of $x$. It also follows from (34b) and (34c) that $B_z^{(1)} = \varepsilon \Psi_1$. Thus, in the single-fluid MHD this field component is separated from two other equations in (34), and, therefore, does not affect the process of forced reconnection. Finally, two equations for the functions $\psi(x,t)$ and $\phi(x,t)$ take the form

$$\frac{\partial \psi}{\partial t} = \varepsilon kx\phi + \frac{1}{S}\psi'', \qquad (35a)$$

$$\frac{\partial}{\partial t}(\phi'') = -\varepsilon k x \psi'' \qquad (35b)$$

Initially, when the CS thickness is not yet sufficiently small (see below), plasma resistivity plays no role, so the process can be described in terms of the ideal MHD [i.e. without the last term on the r.h.s. of Eq.(35a)]. Thus, if $\Delta(t) \equiv \Delta x/l$ is a non-dimensional thickness of the CS, then at $x \leq \Delta$ the magnetic field component $B_y^{(1)}$, jump of which across the CS is equal, according to (19), to $\{B_y\} \sim \varepsilon B_0 a/l$ ( hence, in the adopted non-dimensional units, $\{B_y^{(1)}\} \sim 1$), can be approximated as $B_y^{(1)} \approx x/\Delta$. Therefore, the respective flux-function, $\psi_i$ (index "i" indicates here the ideal MHD description) takes there the form

$$\psi_i(x,t) \sim x^2/\Delta(t) \qquad (36)$$

It means that its characteristic value, i.e., $\psi_i(x \sim \Delta) \sim \Delta$, and one can make the following estimates:

$$\frac{\partial \psi_i}{\partial t} \sim \frac{d\Delta}{dt}, \quad \psi'' \sim \frac{1}{\Delta} \qquad (37)$$

Then, since $\phi'' \sim \phi/\Delta^2$, it follows from (35b) that $\frac{\partial}{\partial t}\left(\frac{\phi}{\Delta^2}\right) \sim k$, so $\phi \sim kt\Delta^2$. By inserting it into Eq. (35a) and using relations (37), one finally gets $(d\Delta/dt) \sim k\Delta\phi \sim k^2 t \Delta^3$, a simple integration of which yields

$$\Delta(t) \sim (kt)^{-1} \qquad (38)$$

Thus, in the absence of resistivity, such shrinking of the CS would bring about, though only asymptotically in time, the singular ideal MHD equilibrium (18). However, this process comes to the end when, eventually, a finite plasma resistivity intervene. It happens when the resistive term in Eq.(35a), $S^{-1}\psi''$, becomes significant, and the respective time-scale can be estimated as follows. According to (37) and (38), in Eq.(35a) the term $(\partial \psi/\partial t) \sim (d\Delta/dt) \sim 1/kt^2$, while $S^{-1}\psi'' \sim S^{-1}\Delta^{-1} \sim S^{-1}kt$. Therefore, the two become comparable at

$$t \sim t_1 \sim k^{-2/3} S^{1/3} \qquad (39)$$

The subsequent, at $t > t_1$, resistive evolution of the CS proceeds in the so-called "constant-$\psi$" regime (Furth et al 1963). Indeed, although during the ideal MHD phase of the CS evolution, at $t < t_1$, the effect of resistivity is weak, the resistive term in (35a) is not exactly equal to zero, hence, some degree of magnetic reconnection does take place. The amount of reconnected magnetic flux, $\psi_r$ is equal to $\psi(x=0)$ and, according to (37) and (38), it can be estimated as follows:

$\frac{d\psi(0)}{dt} = \frac{1}{S}\psi'' \sim S^{-1}\Delta^{-1} \sim S^{-1}kt$, hence at $t \leq t_1$ the reconnected flux accumulates as

$$\psi_r(t) = \psi(0) \sim S^{-1}kt^2 \qquad (40)$$

On the other hand, CS shrinking leads to reduction of its internal magnetic flux as $\Delta\psi = [\psi_i(x \sim \Delta) - \psi_i(0)] \sim \Delta \sim (kt)^{-1}$. Therefore, at $t \sim t_1$ one gets $(\Delta\psi) \sim \psi(0)$, and the further reconnection makes $(\Delta\psi) < \psi(0)$. This validates the "constant-$\psi$" approximation in analysing the CS dynamics at $t > t_1$. Since the relevant time-scale is long compared with the Alfven transit time $\tau_A$, outside the CS the system remains in a state of the quasistatic evolution. Therefore, the flux-function perturbation there (the external solution, $\psi_{ext}$) can be expressed as a superposition of the equilibrium solutions (14) and (18):

$$\psi_{ext}(x,t) = f(t)\psi^{(r)}(x) + [1-f(t)]\psi^{(i)}(x) \qquad (41)$$

(note that such a form is necessary to preserve the boundary conditions at $x = x_b^{(\pm)}$). Physical meaning of the function $f(t)$, which is a fraction of completed reconnection, becomes evident after matching (41) with the solution inside the CS, where at $t > t_1$ $\psi(x,t) \approx \psi_r(t)$. Thus, bearing in mind that $\psi^{(i)}(0) = 0$, one gets $f(t) = \psi_r(t)/\psi^{(r)}(0)$, where $\psi^{(r)}(0) \sim 1$ [this is in the adopted non-dimensional units, see Eq.(14)] is the terminal value of the reconnected flux. It is reached when the process of forced reconnection comes to the end, and the regular equilibrium, $\psi^{(r)}$, becomes fully established. Note that, according to Eq.(40), $\psi_r(t \sim t_1) \sim S^{-1}kt_1^2 \sim S^{-1/3}k^{-1/3}$, which formally could yield $\psi_r(t_1) > 1$ [as well as $\Delta(t_1) > 1$, see Eq.(37)] for $k < S^{-1}$. However, one should remember that in this case the very concept of forced reconnection makes no sense. Indeed, such a process is of interest only if its time-scale is much shorter than the global resistive time $\tau_\eta = l^2/\eta$, which in non-dimensional units translates into $l^2/\eta\tau_A = S$. Therefore, it requires that $t_1 \sim k^{-2/3}S^{1/3} < S$, i.e. $k > S^{-1}$, so that both $\Delta(t_1)$ and $\psi_r(t_1)$ are small.

At the resistive stage of the process, i.e. for $t > t_1$, temporal evolution of the reconnected flux, $\psi_r(t)$, and the CS thickness, $\Delta(t)$, can be obtained in the following way. Firstly, as long as the reconnected magnetic flux is still small, i.e. $f(t) \sim \psi_r(t) \ll 1$, the ideal equilibrium dominates in the external solution (41). Therefore, the discontinuity of $B_y^{(1)}$ across the CS persists, hence, as it was at the ideal MHD phase, the second of relations (37) holds: $\psi'' \cdot \Delta \sim 1$. Secondly, in this regime the convective and resistive terms on the r.h.s. of Eq.(35a) should be comparable, which yields $S^{-1}\psi'' \sim S^{-1}\Delta^{-1} \sim k\Delta\phi$, hence $\phi \sim S^{-1}k^{-1}\Delta^{-2}$. Then, by inserting this expression for $\phi$ into Eq.(35b), one can derive how the CS thickness varies in time. Thus, $\partial(\phi'')/\partial t \sim \partial(\phi/\Delta^2)/\partial t \sim \partial(S^{-1}k^{-1}\Delta^{-4})/\partial t \sim kx\psi'' \sim k$, which yields

$$\Delta(t) \sim 1/S^{1/4}k^{1/2}t^{1/4} \qquad (42)$$

As seen from (42), shrinking of the CS continues, which, as follows from (35a), results in the reconnection rate $(d\psi_r/dt) \sim S^{-1}\psi'' \sim S^{-1}\Delta^{-1} \sim S^{-3/4}k^{1/2}t^{1/4}$, hence

$$\psi_r(t) \sim S^{-3/4}k^{1/2}t^{5/4} \qquad (43)$$

Such a regime proceeds until the reconnected flux becomes of order of unity, which, according to (43), occurs at $t \sim \tau_r$, where

$$\tau_r \sim S^{3/5}k^{-2/5} \qquad (44)$$

The point is that that at this time the reconnection fraction factor introduced in Eq. (41), $f(t) \sim \psi_r(t)$, becomes already not small. Therefore, the strength of the CS, which is determined by the amplitude of the ideal MHD solution there, starts to diminish. Although this reduces the pace of reconnection, the reconnection factor $f$ becomes very close to unity at the time-scale of several $\tau_r$ (Hahm & Kulsrud 1985). Thus, Eq. (44) presents the time-scale of forced magnetic reconnection in the framework of the inviscid single fluid MHD theory.

Such an analysis can be extended to the case of the visco-resistive MHD, when viscous force $\rho\nu\nabla^2\vec{V}$ is added to the r.h.s. of the equation of motion (31) ($\nu$ is kinematic plasma viscosity). The effect of viscosity, which, quite obviously, slows down the process of forced reconnection, becomes significant when the respective Prandtl number $P \equiv \nu/\eta$ is large enough, namely $P \gg 1$ (Fitzpatrick 2003) (which is assumed in what follows). A straightforward derivation shows that inclusion of viscosity yields the following modification of Eq.(35b):

$$\frac{\partial}{\partial t}(\phi'') = -\varepsilon k x \psi'' + \frac{P}{S}\frac{d^2}{d^2 x}(\phi'') \qquad (45)$$

As a result, the Ideal MHD phase, which corresponds to $t < t_1 \sim k^{-2/3} S^{1/3}$ in the case of the inviscid plasma, becomes divided into two sub-phases. Initially, when the viscous force is still weak, the CS shrinks in the same manner as discussed above for the inviscid plasma [see Eqs. (37) and (38)]. At this sub-phase the stream-function evolves as $\phi \sim kt\Delta^2 \sim (kt)^{-1}$, so the inertial term in Eq.(45) remains constant, $\partial \phi''/\partial t \sim k$, while the viscous term is increasing as $(P/S)d^2(\phi'')/d^2 x \sim (P/S)\phi/\Delta^4 \sim (P/S)k^3 t^3$. Therefore, the two terms become comparable at $t \sim t_1^{(v)} \sim k^{-2/3}(P/S)^{-1/3} \sim t_1/P^{1/3} \ll t_1$, so at $t > t_1^{(v)}$ the plasma inertia plays no role, and in the equation of motion (45) the magnetic force is balanced by the viscous one, hence, $(P/S)\phi/\Delta^4 \sim \varepsilon k x \psi'' \sim k$, i.e.

$$\phi \sim (P/S)^{-1} k \Delta^4. \qquad (46)$$

By inserting this $\phi$ into Eq. (35a) (still no resistivity, the ideal MHD!), and recalling from (37) that $\partial \psi / \partial t \sim d\Delta/dt$, one gets $d\Delta/dt \sim k\Delta\phi \sim (P/S)^{-1} k \Delta^5$, hence,

$$\Delta(t) \sim \frac{(P/S)^{1/4}}{k^{1/2} t^{1/4}} \qquad (47)$$

Such viscous dominated shrinking of the CS proceeds until the plasma resistivity comes into play at some time $t \sim t_2^{(v)}$. This time can be estimated by equating the advective and resistive terms on the r.h.s. of Eq.(35a): $\varepsilon k x \phi \sim k \Delta \phi \sim (P/S)^{-1} k^2 \Delta^5$ and $S^{-1}\psi'' \sim S^{-1}\Delta^{-1}$. With help of (47), it yields $t_2^{(v)} \sim (P/S)^{1/3} k^{-2/3} S^{2/3} \sim t_1 P^{1/3} \gg t_1^{(v)}$.

Similar to the case of the inviscid plasma, at $t > t_2^{(v)}$ the finite resistivity brings about the "constant-$\psi$" regime of forced magnetic reconnection. Its time-scale, $\tau_r^{(v)}$, can be obtained in the following way. In this regime the advective and resistive terms in Eq.(35a) should be of the same order of magnitude. Therefore, by using (46), their comparison allows one to estimate the CS thickness. Thus, $\varepsilon k x \phi \sim k \Delta \phi \sim (P/S)^{-1} k^2 \Delta^5$, while $S^{-1} \psi'' \sim S^{-1} \Delta^{-1}$, hence $\Delta \sim (P/S)^{1/6} k^{-1/3} S^{-1/6}$. This yields the reconnection rate $d\psi_r / dt \sim S^{-1} \Delta^{-1} \sim k^{1/3} S^{-1/6} (P/S)^{1/6}$, hence, $\psi_r(t) \sim k^{1/3} S^{-1/6} (P/S)^{-1/6} t$. Then, since $\psi_r(t \sim \tau_r^{(v)})$ should be of order of unity, the respective reconnection time is equal to

$$\tau_r^{(v)} \sim S^{2/3} P^{1/6} k^{-1/3} \qquad (48)$$

Note, however, that by the very meaning of the forced reconnection problem the above time-scale should be short compared to the global resistive time $\tau_\eta$, which in the adopted non-dimensional units is equal to $\tau_\eta = S$. This requirement puts an upper limit on the allowed value of the Prandtl number $P$, namely $P \ll S^2 k^2$ (which also ensures that $\tau_r^{(v)} \gg t_2^{(v)}$).

So far we considered a single, one-off event of forced reconnection, which results in release of the excess magnetic energy given by Eq.(23). However, in many applications, first of all in plasma astrophysics, the interest is in a continuous plasma heating provided by an ongoing magnetic reconnection. For example, such a situation takes place in the solar corona, where shuffling of photospheric footpoints of magnetic field lines supply the energy flux sufficient to maintain very high temperature of the coronal plasma (see, e.g. Golub & Pasachoff 2010; Vekstein 2016). Therefore, in order to model such a situation in the framework of a simple model of forced magnetic reconnection, we now assume that some external source provides periodic in time deformation of the external boundary surfaces, so that the perturbation amplitude $a$ in (8) varies with time as $a(t) = a_0 \exp(-i\omega t)$. In the context of magnetic reconnection the interest is in a quasi-static perturbation, when $\omega \tau_A \ll 1$. Therefore, the system should remain close to a force-free magnetic equilibrium, hence, the respected external (outside of the CS) flux-function can be represented, similar to (41), as a superposition of the two equilibria as

$$\psi_{ext}(x,t) = [f\psi^{(r)}(x) + (1-f)\psi^{(i)}(x)]\exp(-i\omega t), \qquad (49)$$

The amplitude $f$, which now depends on the perturbation frequency $\omega$, has to be obtained by considering plasma dynamics inside the CS (the internal solution). However, some general conclusions about it can be made *a priori* from the following physical argument. Thus, at low frequency, when $\omega\tau_r \ll 1$, the reconnection is fast enough to bring the system close to the reconnected equilibrium $\psi^{(r)}$, hence in this limit $f \to 1$. In the opposite case, $\omega\tau_r \gg 1$, when the reconnection time is long compared to the time-scale of external driving, one may expect the system to remain close to the ideal MHD equilibrium, i.e. $f \to 0$.

In the case of periodic boundary perturbation Eqs.(35) take the form

$$-i\tilde{\omega}\psi = \varepsilon k x\phi + S^{-1}\psi'', \quad -i\tilde{\omega}\phi'' = -\varepsilon kx\psi'', \qquad (50)$$

where $\tilde{\omega} \equiv \omega\tau_A$ is a non-dimensional driving frequency. By adopting the "constant-$\psi$" approximation (its applicability is discussed below), one can put $\psi(x) \approx \psi(0)$, and reduce then (50) to a single equation for the stream-function $\phi(x)$:

$$S^{-1}\phi'' - i\frac{(\varepsilon kx)^2}{\tilde{\omega}}\phi + \varepsilon kx\psi(0) = 0 \qquad (51)$$

Furthermore, since (51) contains two small parameters: $\tilde{\omega} \ll 1, S \ll 1$, it is convenient to re-scale all variable in such a way that the resulting equation becomes parameter-free. Thus, after substitutions

$$x = (\tilde{\omega}/\varepsilon^2 k^2 S)^{1/4}\xi, \quad \phi = \psi(0)(\tilde{\omega}^3 S/\varepsilon^2 k^2)^{1/4}\chi(\xi), \quad (52)$$

Eq.(51) transforms into a "standard" form equation for the function $\chi(\xi)$:

$$\chi'' - i\xi^2\chi + \xi = 0 \qquad (53)$$

Our interest is in its odd solution tending to zero at $\xi \to \infty$, and since Eq.(53) is parameter-free, one can conclude that this solution is of order of unity, $\chi \sim 1$, and it varies on the scale of $\Delta\xi \sim 1$. Therefore, the re-scaling (52) indicates that the CS thickness, $\Delta$, is of order of

$$\Delta \sim (\tilde{\omega}/S)^{1/4} \ll 1, \qquad (54)$$

while the stream-function inside the CS is of order of

$$\phi \sim \psi(0)(\tilde{\omega}^3 S)^{1/4} \qquad (55)$$

(for simplicity, it is assumed here that $\varepsilon \sim 1, k \sim 1$). Now one can estimate the magnitude of electric current, i. e. $\psi''$, there. Thus, as follows from (50), $\psi'' \sim \tilde{\omega}\phi''/\Delta \sim \tilde{\omega}\phi/\Delta^3$, which with help of Eqs. (54) and (55), yields

$$\psi'' \sim \psi(0)\tilde{\omega}S \qquad (56)$$

This quantity determines the discontinuity of $B_y^{(1)}$, i. e. the jump of $\psi'$ across the CS: $\{\psi'\} = \int \psi'' dx \sim \psi''\Delta \sim \psi(0)\tilde{\omega}^{5/4}S^{3/4}$, which in terms of the tearing mode parameter $\Delta'$ [see Eq.(30)] reads

$$(\Delta')_{int} = \{\psi'\}/\psi(0) \sim \tilde{\omega}^{5/4}S^{3/4} = (\omega\tau_A)^{5/4}S^{3/4} \sim (\omega\tau_r)^{5/4}, \qquad (57)$$

where $\tau_r \sim \tau_A S^{3/5}$ is the time-scale of forced magnetic reconnection [see Eq.(44)]. Then, in order to derive the amplitude $f$ in the external solution (49), one should match expression (57) with the respective value of $\Delta'$ that follows from (49). The latter yields $\{\psi'_{ext}\} = (1-f)\{(\psi^{(i)})'\} \sim (1-f), \psi_{ext}(0) = f\psi^{(r)}(0) \sim f$ (see Eqs. (14) and (18) for $\psi^{(r)}$ and $\psi^{(i)}$), hence $(\Delta')_{ext} \sim (1-f)/f$. Comparing it with (57), one gets

$$f \approx \frac{1}{1+(\omega\tau_r)^{5/4}}, \qquad (58)$$

which confirms the beforehand expectation that $f \to 1$ when $\omega\tau_r \to 0$, and $f \to 0$ if $\omega\tau_r \to \infty$.

A rigorous solution of Eq.(53) (Vekstein & Jain 1999) yields

$$f = f_1 + if_2; \quad f_1 = \frac{(a^2+b^2)+aZ}{(a+Z)^2+b^2}; \quad f_2 = \frac{bZ}{(a+Z)^2+b^2}; \qquad (59)$$

where $Z \equiv (a^2+b^2)(\omega\tau_r)^{5/4}$ and $a+ib = \frac{2\pi\Gamma(3/4)}{\Gamma(1/4)}\exp(i3\pi/8)$

As seen from (59), the amplitude $f$ is a complex number, which indicates a phase shift, or, in physical terms, a temporal lag between the external driver and the internal response of the system. It is this lag that determines the energy dissipation rate caused by forced reconnection. Indeed, since all parameters of the system vary in time periodically, the sought after energy dissipation rate, $Q_r$, can be derived as a mean power of the external force which provides for the continuous boundary deformation. Thus, similarly to (20), one gets

$$Q_r = -2\left\langle\left\langle \text{Re}\left[\frac{B^2(x_b^{(+)})}{8\pi}\right]\cdot \text{Re}\left[\frac{dx_b^{(+)}}{dt}\right]\right\rangle\right\rangle,$$

where symbol $\langle\langle\rangle\rangle$ means averaging over the variation along the y-coordinate and over the temporal period of oscillations. Then, a straightforward calculation yields the following result:

$$Q_r = \frac{\Delta W_r}{\tau_r}F(\omega\tau_r), \quad F(\omega\tau_r) = (\omega\tau_r)f_2(\omega\tau_r), \quad (60)$$

where $\Delta W_r$ is the excess magnetic energy given by Eq.(23), and $f_2$ is the imaginary part of the amplitude $f$ in (59). The relaxation function $F(\omega\tau_r)$ is plotted in Fig.3. At $(\omega\tau_r)\ll 1$ one gets from (57) that $F\propto(\omega\tau_r)^{9/4}$, while for $(\omega\tau_r)\gg 1$ it yields $F\propto(\omega\tau_r)^{-1/4}$.

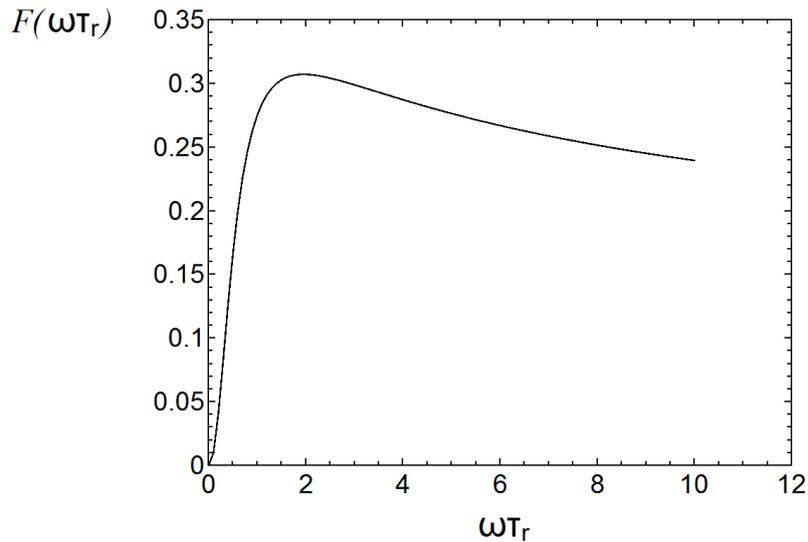

Figure 3. Reconnective relaxation function $F(\omega\tau_r)$.

This exhibits features typical for a relaxation process, with the dissipation becoming most effective when the time-scale of an external driving is comparable to the characteristic time of an internal relaxation. Another example is the well-known effect of "second viscosity" in a gas or liquid (Landau & Lifshitz 1987), which leads to frequency-dependent enhanced attenuation of sound waves. In the context of forced reconnection, the explanation is quite simple. If the time-scale $\tau_r$ is too long, the reconnection process is not effective enough. On the other hand, if the reconnection is too fast, it does not allow build-up of the excess magnetic energy by bringing the magnetic field (49) close to the relaxed state with $f \to 1$.

Clearly, the actual physical mechanism responsible for the dissipation power (60) is the Joule plasma heating inside the CS. Therefore, it could be instructive to demonstrate by direct calculation that the Joule heating rate, $Q_J$, matches the $Q_r$ in (60) (as it should be due to the energy conservation law). Thus, the sought-after $Q_J$ can be estimated as $Q_J \sim \dfrac{j_{1z}^2}{\sigma}(\Delta x)$, where $\sigma$ is plasma conductivity, and $(\Delta x)$ is the CS thickness. Then, since $j_{1z} \approx \dfrac{c}{4\pi}\dfrac{\partial^2 \Psi_1}{\partial^2 x}$, by using non-dimensional units introduced above, one gets

$$Q_J \sim \frac{c^2}{4\pi\sigma} \cdot \frac{B_0^2 a^2}{8\pi d^3}(\psi'')^2 \Delta \sim \frac{1}{S\tau_A} \cdot \frac{B_0^2 a^2}{8\pi d}(\psi'')^2 \Delta ,$$

with $\psi''$ and $\Delta$ given by Eqs.(56) and (54). Furthermore, since in the non-dimensional units $\psi^{(r)}(0) \sim 1$ and $\psi^{(i)}(0) = 0$, it follows from (49) that $\psi(0) \sim f$. Finally, all these, together with (58), yield the following expression for the Joule heating rate:

$$Q_J \sim \frac{B_0^2 a^2}{8\pi d} \cdot \frac{(\omega\tau_A)^{9/4} S^{3/4}}{\tau_A[1+(\omega\tau_r)^{5/4}]^2} \sim \frac{B_0^2 a^2}{8\pi d} \cdot \frac{(\omega\tau_r)^{9/4}}{\tau_r[1+(\omega\tau_r)^{5/4}]^2} , \qquad (61)$$

(recall that, according to (44), $\tau_A \sim \tau_r S^{-3/5}$). As seen from (61), this expression is in full qualitative agreement with (60), providing the correct scaling of the energy dissipation rate in both limits of slow, $(\omega\tau_r \ll 1)$, and fast, $(\omega\tau_r \gg 1)$, external driving.

Consider now applicability of the adopted so far "constant-$\psi$" approximation. Proceeding first in a formal way, this issue can be resolved as follows. Since $\psi$ is an even function of $x$, its variation inside the CS, $\Delta\psi$, can be estimated as $\Delta\psi \sim \psi''\Delta^2$, hence, with help of (54) and (56), one gets $\Delta\psi \sim \psi(0)(\omega\tau_A)^{3/2} S^{1/2}$. This variation exceeds $\psi(0)$ at $\omega > \omega_1 \sim \tau_A^{-1} S^{-1/3}$, hence, invalidating the "constant-$\psi$" assumption. Therefore, the relaxation function $F(\omega\tau_r)$ shown in Fig.3 makes sense only for $\omega < \omega_1$. Note, however, that since $\omega_1\tau_r \sim S^{4/15} \gg 1$, only a far "tail" of this figure is affected, while all conclusions about the energy dissipation due to the internal relaxation process remain intact.

The very same conclusion can be reached from a more physical viewpoint by comparing the CS skin-time $\tau_s = (\Delta x)^2/\eta = l^2\Delta^2/\eta \sim \tau_A(\omega\tau_A)^{1/2} S^{1/2}$ [see Eq.(54)] with the driving time-scale $\Delta t \sim \omega^{-1}$. As seen, for $\omega > \omega_1$ one gets $\tau_s > \Delta t$, i.e. finite plasma resistivity has no enough time to level-off the flux function inside the CS, which is necessary for validity of the "constant-$\psi$" condition. Moreover, this can be viewed as a hint that for $\omega > \omega_1$ the finite plasma resistivity does not play a role at all, hence, the response of the system to such a perturbation can be studied in the framework of the ideal MHD. Note, that $\omega > \omega_1$ corresponds to the characteristic time-scale of $t < \omega_1^{-1} \sim \tau_A S^{1/3}$, which is nothing else than the time $t_1$ introduced above [see Eq.(39)] while considering the one-off event of forced magnetic reconnection. It has been demonstrated there that the plasma resistivity is indeed insignificant at $t < t_1$.

It does not mean, however, that there is no energy dissipation in the realm of the ideal MHD, when the driving frequency exceeds $\omega_1$. In this case absorption of the energy supplied by external driver is associated with the **Alfven resonances** (Chen & Hasegawa 1974). Their spatial location, $x_A^{(r)}$, is determined by the condition $\omega = \pm\vec{k} \cdot \vec{V}_A(x_A^{(r)})$, where $\vec{V}_A(x) = \vec{B}^{(0)}(x)/4\pi\rho$. Therefore, in the case of the initial magnetic field (7), the wave-vector $\vec{k}$ directed along the y-axis, and a quasi-static driving: $\omega\tau_A \ll 1$, one gets $x_A^{(r)} \approx \pm l(\omega\tau_A)$ (it is assumed, as before, that $\alpha l \sim 1, kl \sim 1$). Thus, from the viewpoint of Alfven resonances the transition from the ideal MHD dissipation to the resistive one caused by magnetic reconnection can be explained as follows. The thickness, $\Delta x$, of the CS located at $x = 0$ increases with the plasma resistivity (which means reduction of $S$) according to Eq.(54). Therefore, at $\omega < \omega_1$, the obtained above $x_A^{(r)}$ becomes, formally speaking, smaller than $\Delta x$. As the result, Alfven resonances cease to exist, being now "buried" inside the resistive CS. A combined diagram illustrating both the resonant and the reconnective dissipation power is plotted in Fig.4 (Vekstein 2000). It clearly exhibits a characteristic relaxation "hump" along the otherwise monotonically decreasing rate of the magnetic energy dissipation.

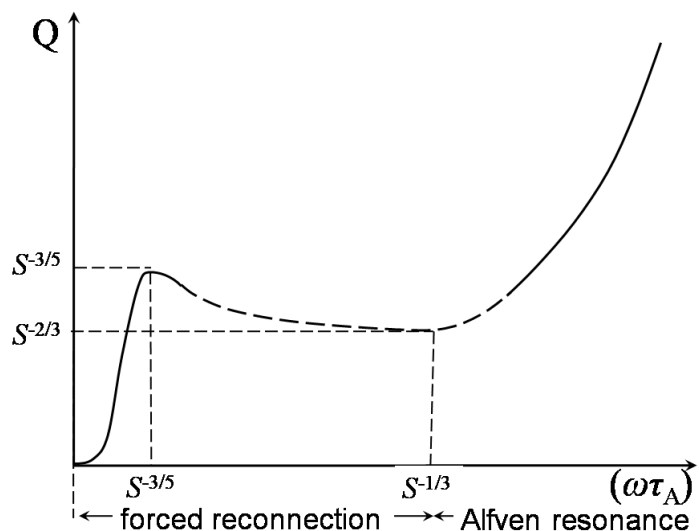

Figure 4. Combined diagram of the reconnective and resonant magnetic energy dissipation power.

## 5. Hall-mediated forced magnetic reconnection

According to results discussed in Section 4, thickness $\Delta x$ of the resistive CS, which plays a central role in the process of forced reconnection, scales as $\Delta x \sim lS^{-2/5}$ [see Eqs.(42) and (44)]. Therefore, in many practical applications with a very large value of the Lundquist number $S$, $\Delta x$ becomes so small that the single-fluid MHD description is not applicable. In this case flows of electrons and ions inside the CS are strongly decoupled, and since magnetic field is advected by the flow of electrons, the lighter plasma component , the bulk plasma velocity $\vec{V}$ in the magnetic induction equation (1) should be replaced by the velocity of the flow of electrons, $\vec{V}_e$. Since the latter is equal to $\vec{V}_e = \vec{V} - \vec{j}/ne = \vec{V} - c(\vec{\nabla} \times \vec{B})/4\pi ne$, Eq.(1) takes the form

$$\frac{\partial \vec{B}}{\partial t} = \vec{\nabla} \times (\vec{V} \times \vec{B}) - \frac{c}{4\pi n e} \vec{\nabla} \times [(\vec{\nabla} \times \vec{B}) \times \vec{B}] + \eta \nabla^2 \vec{B} \qquad (62)$$

The second term on the r.h.s. of (62) accounts for what is called **Hall effect**. One can easily verify that its role becomes significant when spatial scale of the magnetic field variation, $L$, is comparable to the ion inertial length $d_i \equiv c/\omega_{pi} = c/\sqrt{4\pi n e^2/m_i}$. Indeed, since in a low-$\beta$ plasma the characteristic dynamical velocity is Alfven velocity $V_A = B/\sqrt{4\pi n m_i}$, the first term on the r.h.s. of (62) can be estimated as $\vec{\nabla} \times (\vec{V} \times \vec{B}) \sim V_A B/L$, while the Hall term is of order of $\frac{c}{4\pi n e} \vec{\nabla} \times [(\vec{\nabla} \times \vec{B}) \times \vec{B}] \sim \frac{c}{4\pi n e} B^2/L^2$. Thus, for their ratio one gets $\frac{c}{4\pi n e} \cdot \frac{B}{V_A L} = d_i/L$.

As far as magnetic reconnection is concerned, a major difference caused by the Hall effect is appearance of the quadrupole structure (see below) in the out-of-plane magnetic field component $B_z^{(1)}$ (see, e.g. Ren et al. 2005). Furthermore, the Hall effect also results in the plasma flow in this direction, $V_z$, which was absent in the case of the single-fluid MHD discussed in Section 4. Therefore, the set of governing equations (33) should be complemented by the equation of motion for $V_z$:

$$\frac{\partial V_z}{\partial t} = \frac{1}{4\pi\rho} \left( \frac{dB_z^{(0)}}{dx} \cdot \frac{\partial \Psi_1}{\partial y} - \frac{d\Psi_0}{dx} \cdot \frac{\partial B_z^{(1)}}{\partial y} \right) \qquad (63)$$

(note that, as follows from Eqs. (34a,b) the r.h.s. of (63) indeed vanish in the single-fluid MHD case). After that the linearized form of Eq. (62) yields the following evolution equations for perturbations $\Psi_1$ and $B_z^{(1)}$ that now replace Eqs. (33a,b):

$$\frac{\partial \Psi_1}{\partial t} = -\frac{d\Psi_0}{dx} \cdot \frac{\partial \Phi}{\partial y} + \eta \nabla^2 \Psi - \frac{c}{4\pi n e} \left( \frac{dB_z^{(0)}}{dx} \cdot \frac{\partial \Psi_1}{\partial y} - \frac{d\Psi_0}{dx} \cdot \frac{\partial B_z^{(1)}}{\partial y} \right), \qquad (64a)$$

$$\frac{\partial B_z^{(1)}}{\partial t} = -\frac{dB_z^{(0)}}{dx} \cdot \frac{\partial \Phi}{\partial y} + \eta \nabla^2 B_z^{(1)} - \frac{d\Psi_0}{dx} \cdot \frac{\partial V_z}{\partial y} + \frac{c}{4\pi n e} \left( \frac{d^3\Psi_0}{d^3 x} \cdot \frac{\partial \Psi_1}{\partial y} - \frac{d\Psi_0}{dx} \cdot \frac{\partial}{\partial y} \nabla^2 \Psi_1 \right), \quad (64b)$$

while Eq.(33c) for the stream-function $\Phi$ remains unchanged.

Consider now symmetry properties of the perturbations. As far as the flux-function $\Psi_1$ and the stream-function $\Phi$ are concerned, they are the same as in the single-fluid MHD case:

$$\Psi_1(x,y,t) = \psi(x,t)\cos ky, \quad \Phi(x,y,t) = \phi(x,t)\sin ky$$

On the contrary, the magnetic field component $B_z^{(1)}$ is now a superposition of both modes, and it can be written as

$$B_z^{(1)} = \alpha\psi(x,t)\cos ky + b(x,t)\sin ky$$

The last term in this equation, which is entirely due to the Hall effect, represents the above-mentioned quadrupole structure ($b$ is an odd function of $x$), and, as demonstrated below, it is at the heart of the Hall-mediated magnetic reconnection. Finally, as seen from (63), the out-of-plane plasma velocity component has the form $V_z(x,y,t) = v(x,t)\cos ky$, where $v$ is an even function of $x$.

Then, by introducing non-dimensional variables and following the simplification procedure identical to those adopted in Section 4, one arrives to the following set of equations for functions $\psi(x,t), \phi(x,t), b(x,t)$ and $v(x,t)$:

$$\frac{\partial \psi}{\partial t} = \varepsilon kx\phi + \frac{1}{S}\psi'' - \varepsilon kxhb, \qquad (65a)$$

$$\frac{\partial}{\partial t}(\phi'') = -\varepsilon kx\psi''', \qquad (65b)$$

$$\frac{\partial b}{\partial t} = -\varepsilon kxv + \frac{1}{S}b'' - \varepsilon kxh\psi'', \qquad (65c)$$

$$\frac{\partial v}{\partial t} = \varepsilon kxb \qquad (65d)$$

Here Hall effect is accounted for by the last terms in Eqs. (63a,c), both of which are proportional, as already anticipated, to the Hall parameter $h \equiv d_i/l$. Thus, Eqs.(65) reduce to Eqs.(35) by putting $h=0$. In what follows, it is assumed that $h \ll 1$, which is the case for a majority of applications. Nevertheless, as demonstrated below, even under this condition the reconnection process could be Hall-mediated because of a very small thickness of the CS, $\Delta x \ll l$ (or, in the non-dimensional units used in (65), $\Delta \ll 1$).

Then, the initial phase of the process is identical to that in the single-fluid MHD: the ideal MHD shrinking of the CS described by Eq.(38). In order to find out when the Hall effect intervenes, one can first use Eqs.(65c), (36) and (37) for estimating the quadrupole field $b$ generated in the CS during this phase of its evolution: $\frac{\partial b}{\partial t} \approx -\varepsilon khx\psi'' \sim kh$, hence, $b \sim kht$. Therefore, the Hall term in Eq.(65a) is of order of $kh\Delta b \sim kh^2$. On the other hand, the l.h.s. of this equation, $\partial \psi / \partial t$, is of order of $k^{-1}t^{-2}$ [see Eq.(37)], which indicates that the Hall effect comes into play at

$$t \sim t_H \sim k^{-1}h^{-1} \qquad (66)$$

Note that according to Eq.(38), at this time the CS thickness $\Delta(t \sim t_H) \sim h$, which in dimensional units translates into $\Delta x \sim d_i$. However, the above-given consideration ignores the effect of a finite plasma resistivity. Thus, the interplay between the two effects, Hall and resistivity, is determined by the relation between the respective time-scales, $t_H$ give in (66), and $t_1$ derived in Eq.(39) (Vekstein & Kusano 2017). Consider first the case when $t_1 < t_H$, i.e. the resistivity comes first, which implies that

$$h < h_1 \sim k^{-1/3}S^{-1/3} \ll 1 \qquad (67)$$

It turns out that in this case the Hall effect does not play any role at all in the process of forced reconnection. In order to demonstrate this, it is necessary to evaluate how generation of the quadrupole magnetic field $b$ proceeds at $t > t_1$, when, according to Section 4, the system enters the "constant-$\psi$" regime of forced reconnection. One can easily verify that at $t > t_1$ the "generating" Hall term in Eq.(65c) is balanced by the resistive diffusion of $b$, so $\varepsilon kxh\psi'' \sim kh$ should be compared with $b''/S \sim b/S\Delta^2$, which with help of (42) yields $b \sim h(S/t)^{1/2}$. Then, by inserting this expression for $b$ into the Hall term in Eq.(65a), one gets $\varepsilon kxhb \sim k\Delta hb \sim k^{1/2}h^2 S^{1/4}/t^{3/4}$, which at $t > t_1$ is small compared to other terms in this equation. Indeed, its ratio to $d\psi_r / dt$ [see Eq.(43)] reads

$$\frac{h^2 S}{t} \sim \left(\frac{h}{k^{-1/3}S^{-1/3}}\right)^2 \cdot \frac{t_1}{t} \ll 1 \text{ under the condition (67)}.$$

In this context it is worth noting that, contrary to a common wisdom, the Hall effect could remain insignificant even when in a course of reconnection the CS thickness, $\Delta x$, gets smaller than the ion inertial length $d_i$. Indeed, in the problem under discussion, it follows from Eqs. (42) and (44), that (in non-dimensional units) $\Delta(t \sim \tau_r) \sim (Sk)^{-2/5}$, which could be smaller than $h$ even under the constraint (67).

Thus, the Hall-mediated regime of forced reconnection requires (as shown below, at least) the inequality opposite to (67):

$$h > h_1 \sim (Sk)^{-1/3}, \qquad (68)$$

which implies that $t_H < t_1$, so the Hall effect comes into play before the plasma resistivity intervenes. Therefore, in this case at $t > t_H$ the system enters a phase of the ideal Hall-MHD, at which the evolution of $\psi_i$ and $b$ is governed entirely by the Hall terms in Eqs.(65a,c). The result is a further shrinking of the CS, which can be derived in the same way as explored in Eqs.(36-38). Thus, it follows now from Eq.(65c) that $\partial b / \partial t \sim kh\Delta \psi'' \sim kh \sim t_H^{-1}$, hence,

$$b \sim t / t_H \qquad (69)$$

By inserting it into (65a), one gets $\partial \psi_i / \partial t \sim khxb \sim \Delta t / t_H^2$, and since, according to (37), $\partial \psi_i / \partial t \sim d\Delta / dt$, it yields

$$\Delta(t) \sim h \exp(-t^2 / t_H^2) \qquad (70)$$

Such exponential shrinking of the CS, which is much faster than that in the case of the single-fluid MHD [see Eq.(38)] originates from the dispersive character of the Hall-MHD waves, whistlers (see, e.g. Bulanov et al 1992).

This, ideal phase of evolution holds until the resistivity eventually intervenes at some time $t \sim t_*$, when thickness of the CS becomes sufficiently small: $\Delta(t_*) << h$. This instant can be obtained by equating the resistive and Hall terms in Eq. (65a) at $t \sim t_*$ as follows. The resistive term is $S^{-1}\psi'' \sim S^{-1}\Delta^{-1}(t)$, while the Hall one is of order of $kh\Delta(t)b(t)$, hence, it yields

$$kh\Delta(t_*)b(t_*) \sim S^{-1}\Delta^{-1}(t_*) \qquad (71)$$

Then, since temporal variation of $\Delta$ is, according to (70), much stronger than that of $b$ in (69), with a logarithmic accuracy the sought after time is $t_* \sim t_H$. Therefore, one can put $b(t_*) \sim 1$ in Eq.(71), which then yields

$$\Delta(t_*) \equiv \Delta_H \sim S^{-1/2}(kh)^{-1/2} \sim S^{-1/2} t_H^{1/2} \qquad (72)$$

Note that the anticipated inequality, $\Delta_H \ll h$, is satisfied indeed because of the condition (68).

The subsequent resistive Hall-MHD regime is quite similar to that in the single-fluid MHD discussed in Section. The only difference is that in the Hall-MHD case advection of the poloidal magnetic field into the reconnecting CS is provided by the Hall effect [the last term in Eq.(65a)] rather than by the plasma inflow [the first term on the r.h.s. of (65a)]. Firstly, one can verify that at $t > t_*$ the reconnection proceeds in the "constant-$\psi$" regime. Indeed, in the course of the CS shrinking its internal magnetic flux is decreasing with time [see Eq.(36)]as $\Delta \psi \sim \Delta$, hence $\Delta \psi(t_*) \sim \Delta_H$. On the other hand, the reconnected magnetic flux $\psi_r = \psi(x=0)$ is increasing with time as $d\psi_r/dt = S^{-1}\psi'' \sim (S\Delta_H)^{-1}$. Thus, $\psi_r(t_*) \sim t_H/S\Delta_H$, which is equal to $\Delta\psi(t_*)$ under $\Delta_H$ given in Eq.(72). Therefore, at $t > t_*$ the variation of $\psi$ inside the CS is small: $\Delta\psi \ll \psi(x) \approx \psi_r$, and the set of equations governing temporal evolution of $\psi_r, \Delta$ and $b$ takes the following form. Firstly, the required discontinuity of $B_y^{(1)}$ across the CS yields that $\Delta \cdot \psi'' \sim 1$. Secondly, at this reconnection regime the resistive term in Eq. (65a), $S^{-1}\psi'' \sim (S\Delta)^{-1}$, should be of the same order of magnitude as the Hall term there, $\varepsilon khxb \sim kh\Delta b$. Thirdly, in Eq.(65c) the magnitude of the generated quadrupole filed $b$ is determined by the balance of the Hall term, $\varepsilon khx\psi'' \sim kh$, and the resistive one, $S^{-1}b'' \sim b/S\Delta^2$. These yield $\Delta \sim \Delta_H, b \sim 1$, which, according to (65a), result in

$$\frac{d\psi_r}{dt} = \frac{1}{S}\psi'' \sim \frac{1}{S\Delta_H} \sim S^{-1/2}(kh)^{1/2} \Rightarrow \psi_r(t) \sim S^{-1/2}(kh)^{1/2}t \qquad (73)$$

It means that if this reconnection regime proceeds until full completion of the process of forced reconnection, i.e. when $\psi_r \sim 1$, the respective reconnection time that follows from (73) would be equal to

$$\tau_r^{(H)} \sim S^{1/2}(kh)^{-1/2} \qquad (74)$$

Note that the scaling (74) yields the reconnection time that does not depend on the ion mass $m_i$. Indeed, the Lundquist number is proportional to $m_i^{-1/2}$, the Hall parameter $h \propto m_i^{1/2}$, while the imposed unit of time is $\tau_A \propto m_i^{1/2}$. Therefore, this regime of reconnection corresponds to the Electron-MHD limit in the theory of forced magnetic reconnection (Bulanov et al 1992).

However, it turns out that such a regime is actually realized only when the Hall parameter $h$ exceeds a certain second threshold, $h_2$, (see below), which is much higher than the first one, $h_1$, given in Eq.(67). Otherwise, at some time $\tilde{t} \ll \tau_r^{(H)}$, the Hall regime (73) gives way to the standard single-fluid MHD reconnection, making the overall reconnection time equal to $\tau_r$ defined in Eq.(44). The reason lies in the double-layer structure of the internal solution during the resistive phase of the Hall-MHD reconnection (Terasawa 1983). Thus, the resistive region, $x \leq \Delta_H$, is surrounded by a much wider layer, $\Delta_H < x < x_H$, where the plasma resistivity plays no role, but the poloidal magnetic field described by the flux-function $\psi$ is still advected towards the reconnection site by the Hall effect [the last term on the r.h.s. of Eq.(65a)]. Therefore, by using Eqs.(65a) and (73), one can evaluate the required level of the quadrupole field $b$ that provides for such advection:

$$kxhb \sim \frac{\partial \psi}{\partial t} \sim S^{-1/2}(kh)^{1/2} \Rightarrow b \sim \frac{S^{-1/2}(kh)^{-1/2}}{x} \sim \frac{\Delta_H}{x} \qquad (75)$$

On the other hand, the very same field (75) generates, according to Eq.(65d), the z-component of the plasma velocity, $v$:

$$\frac{\partial v}{\partial t} = \varepsilon kxb \sim k\Delta_H \Rightarrow v \sim k\Delta_H t \qquad (76)$$

Furthermore, inside the ideal sub-layer the Hall term in Eq.(65c) has to be balanced by the first term on its l.h.s., which is due to this velocity component. Hence, by equating these two terms, one can evaluate the electric current there, i.e. $\psi''$, as $kxv \sim kxh\psi''$, which with help of (76) yields $\psi'' \sim kh^{-1}\Delta_H t \sim h^{-2}\Delta_H(t/t_H)$. This current accelerates the poloidal plasma flow at the rate given by Eq.(65b):

$$\frac{\partial}{\partial t}(\phi'') \sim kx\psi'' \sim kxh^{-2}\Delta_H \frac{t}{t_H} \Rightarrow \phi'' \sim h^{-3}\Delta_H x \left(\frac{t}{t_H}\right)^2 \Rightarrow \phi \sim \Delta_H \left(\frac{x}{h}\right)^3 \left(\frac{t}{t_H}\right)^2$$

Finally, by inserting this expression for $\phi$ into Eq.(65a), one can estimate the width of the ideal Hall-MHD sub-layer $x_H$. Indeed, at $x \sim x_H$ advection of the magnetic field by the plasma flow, described by the first term on the l.h.s. of (65a), becomes comparable to that provided by the Hall term. Hence, $\varepsilon k x_H \phi(x \sim x_H) \sim \varepsilon k x_H h b(x \sim x_H)$, which with help of (75) yields

$$x_H \sim h(t_H / t)^{1/2}, \qquad (77)$$

so at $x > x_H$ the Hall effect is not important, and the single-fluid MHD applies.

Therefore, the Hall-MHD regime of reconnection described by Eq.(73) holds as long as the thickness of the resistive sub-layer, $\Delta_H$, is smaller than $x_H$, i.e., according to Eqs. (72) and (77), $t < \tilde{t} \sim Sh^2$. This leaves one with the following two possibilities. If

$$h > h_2 \sim S^{-1/5} k^{-1/5}, \qquad (78)$$

[note that $h_2 >> h_1 \sim S^{-1/3} k^{-1/3}$], one gets $\tilde{t} >> \tau_r^{(H)} \sim S^{1/2} t_H^{1/2}$, hence the Hall-MHD regime has enough time to complete the forced reconnection process. If otherwise, i.e. $h_1 < h < h_2$, a transition from the Hall-MHD regime (73) to the standard single-fluid reconnection (43) occurs at $t \approx \tilde{t}$. At this point the amount of already reconnected magnetic flux is still small: indeed, according to (73), $\psi_r(t \sim \tilde{t}) \sim (h/h_2)^{5/2} << 1$, hence, the main part of reconnection is completed in the single-fluid MHD regime. It is worth emphasizing here that this transition from the Hall- to the single fluid MHD occurs when the thickness of the resistive CS is much smaller than the ion inertial length, $\Delta_H << h$. Moreover, the former reduces even further in the course of the subsequent standard MHD reconnection [see Eq.(42)].

**VI. Nonlinear effects and onset of plasmoid instability.**

A distinctive feature of magnetic reconnection in a system with a large value of the Lundquist number $S$ is formation of a highly elongated CS with a very small thickness, $\Delta x$, which scales as some inverse power of the Lundquist number. Thus, in the case of the single-fluid MHD forced reconnection one gets [see Eqs.(42-44)] $\Delta x \sim l S^{-2/5}$, while in the Hall-MHD regime it scales as $\Delta x \sim l S^{-1/2}$ [see Eq.(72)]. Therefore, such resistive magnetic reconnection is associated with a very high density of the electric current inside the CS, which puts a strong limitation on the applicability of the considered so far linear theory of forced reconnection. Thus, as pointed out by HK, if the amplitude $a$ of the boundary deformation (8) is large enough, namely

$$\delta \equiv \frac{a}{l} > \delta_1 \sim S^{-4/5}, \qquad (79)$$

the linear regime of reconnection (43) gives way to a much slower one, which in the theory of the tearing instability is known as the Rutherford regime (Rutherford 1973). The physical mechanism behind it is the nonlinearity in the torque exerted on the plasma flow [the r.h.s. of Eq.(32c)]. It turns out that under condition (79) the third order (with respect to the parameter $\delta \ll 1$) contribution to this torque balances the linear one which drives the plasma inflow into the reconnection site. This significantly reduces the pace of reconnection, making the time-scale of forced reconnection much longer (see below) than the linear one given in Eq.(44).

Thus, consider now in more detail the torque under question, which, according to (32c), is equal to

$$T = -\frac{\partial \Psi}{\partial x} \cdot \frac{\partial \nabla^2 \Psi}{\partial y} + \frac{\partial \Psi}{\partial y} \cdot \frac{\partial \nabla^2 \Psi}{\partial x} \qquad (80)$$

Therefore, inside the CS the linear torque $T_1 \approx -\frac{d\Psi_0}{dx} \cdot \frac{\partial \nabla^2 \Psi_1}{\partial y}$, is proportional to $\sin ky$, and, according to results of Section 4, it can be estimated as

$$T_1 \sim \delta \cdot \frac{B_0^2}{l^2} \qquad (81)$$

Hereafter we revert to dimensional units and, for the sake of simplicity, assume that the wave-length of the boundary perturbation (8) is comparable to the system size, i.e. $kl \sim 1$, and, as before, the parameter $\varepsilon \equiv \alpha l \sim 1$. This torque drives the flow with the stream-function $\Phi \propto \sin ky$. Since the first-order flux-function $\Psi_1$ is proportional to $\cos ky$, it follows from (80) that the lowest order nonlinear torque able to balance the linear one (hence, also varying along the y-coordinate as $\sin ky$), is the third-order torque

$$T_3 \approx \frac{\partial \Psi_1}{\partial y} \cdot \frac{d^3 \Psi_2^{(0)}}{dx^3}, \qquad (82)$$

where $\Psi_2^{(0)}$ is a part of the second-order flux-function perturbation that does not depend on the y-coordinate. This perturbation originates from the "source" term in Eq.(32a), which is

$$(\vec{\nabla}\Psi_1 \times \vec{\nabla}\Phi)\cdot\vec{z} = \frac{\partial\Psi_1}{\partial x}\cdot\frac{\partial\Phi}{\partial y} - \frac{\partial\Psi_1}{\partial y}\cdot\frac{\partial\Phi}{\partial x} \qquad (83)$$

The two terms in (83), being proportional to $\cos^2 ky$ and $\sin^2 ky$ respectively, generate the required y-independent contribution. It turns out that at $t > t_1 \sim \tau_A S^{1/3}$ [see Eq.(39)], when the system evolves in the resistive constant-$\psi$ regime, the magnitude of $\Psi_{20}$ is determined by the balance between this source term and the resistive term in Eq. (32a). Thus, in order to proceed further, recall, with help of Section 4, that at $t > t_1$ the CS thickness is equal to $\Delta x \sim l S^{-1/4}(t/\tau_A)^{-1/4}$, so at $x \leq \Delta x$

$$\Psi_1 \sim \delta\cdot B_0 l S^{-3/4}(t/\tau_A)^{5/4}, \quad \Phi \sim \delta\cdot V_A l S^{-1/2}(t/\tau_A)^{1/2} \qquad (84)$$

Therefore, the source term (83), where in the constant-$\psi$ phase the first term there is negligible, can be estimated as

$$\frac{\partial\Psi_1}{\partial y}\cdot\frac{\partial\Phi}{\partial x} \sim \frac{\Psi_1}{l}\cdot\frac{\Phi}{\Delta x} \sim \delta^2\cdot B_0 V_A S^{-1}(t/\tau_A)^2, \qquad (85)$$

On the other hand, the respective resistive term in (32a) is of order of

$$\eta\nabla^2\Psi_2^{(0)} \sim \eta\Psi_2^{(0)}/(\Delta x)^2 \sim \frac{\eta}{l^2}S^{1/2}(t/\tau_A)^{1/2}\Psi_2^{(0)} \sim \tau_A^{-1}S^{-1/2}(t/\tau_A)^{1/2}\Psi_2^{(0)},$$

equating which with (85) yields

$$\Psi_2^{(0)} \sim \delta^2\cdot B_0 l S^{-1/2}(t/\tau_A)^{3/2} \qquad (86)$$

(Now one can easily verify that the term $\partial\Psi_2^{(0)}/\partial t$ in Eq.(32a), which has been ignored so far, is small. Thus, according to (86), $\partial\Psi_2^{(0)}/\partial t \sim \delta^2\cdot\tau_A^{-1}B_0 l S^{-1/2}(t/\tau_A)^{1/2}$, hence, its ratio to the source term (85) is of order of $S^{1/2}(t/\tau_A)^{-3/2}$, which is small indeed when $t > t_1 \sim \tau_A S^{1/3}$). Therefore, the third-order torque (82) can be estimated as

$$T_3 \sim \frac{\Psi_1}{l}\cdot\frac{\Psi_2^{(0)}}{(\Delta x)^3} \sim \delta^3\cdot\frac{B_0^2}{l^2}S^{-1/2}\left(\frac{t}{\tau_A}\right)^{7/2}$$

It becomes comparable to the first-order one at

$$t = t_R \sim \tau_A S^{1/7} \delta^{-4/7},  \qquad (87)$$

when the transition to the Rutherford regime of forced reconnection occurs. Of course, such a transition makes sense only when at $t \sim t_R$ the reconnection process is not yet completed, i.e. $t_R < \tau_r$. According to (44) and (87), this is indeed the case under the condition (79).

The important point is that at $t \sim t_R$ the width, $w_r$, of the magnetic islands formed due to the ongoing reconnection [see Eq.(17)] becomes comparable to the CS thickness $\Delta x$ given by Eq.(42). Thus, according to (17), (42) and (43), one gets $w_r(t \sim t_R) \sim l\delta^{1/2} S^{-3/8} (t_R/\tau_A)^{5/8}$, $\Delta(t \sim t_R) \sim lS^{-1/4}(t_R/\tau_A)^{-1/4}$, hence, the two are equal indeed at the transition time $t_R$ given by (87). Then, in the course of the Rutherford regime of forced reconnection, this equality, $\Delta x \sim w_r$ persists, and the rate of reconnection can be estimated as follows. In Eq.(32a) the term associated with the plasma velocity is now insignificant (plasma flow is halted due to the cancellation of the driving torque), therefore, the reconnected magnetic flux, $\Psi_r = \Psi(0)$, evolves in time as

$$d\Psi_r / dt = \eta \nabla^2 \Psi \approx \eta \Psi'' \qquad (88)$$

At the same time the electric current inside the CS remains such as to provide for the required change of the magnetic field across the CS [see Eq.(19)]. Hence, $\{B_y\} \sim B_0 \delta \sim \Psi'' \Delta x$, and since $\Delta x \sim w_r \sim (l/B_0)^{1/2} \Psi_r^{1/2}$, one gets $\Psi'' \sim (B_0 \delta / \Psi_r)^{1/2} (B_0/l)^{1/2}$. Together with (88) it yields the following equation for the normalized reconnected flux $Z(t) \equiv \Psi_r(t)/B_0 l\delta$:

$$Z^{1/2} dZ/dt \sim \delta^{-1/2} \eta/l^2 = \delta^{-1/2}\tau_\eta.$$

Thus, since the completed forced reconnection corresponds to $Z \sim 1$, the time-scale of the Rutherford regime of forced reconnection is equal to

$$\tau_R \sim \tau_\eta \delta^{1/2} \sim \tau_A S \delta^{1/2.} \qquad (89)$$

Therefore, halting of the plasma flow, which supports the reconnection process, makes the reconnection time much longer. It scales now as the first power of the Lundquist number and, therefore, this regime of reconnection falls into the category of the "trivial" reconnection, when the reconnection time is of order of the global resistive time-scale $\tau_\eta$ (see Section 1). The factor of $\delta^{1/2} \ll 1$ in (89) is simply due to the small amount of the reconnected magnetic flux: $\Psi_r \sim B_0 l \delta$.

However, in the above given analysis it was tacitly assumed that the transition to the nonlinear Rutherford regime of forced reconnection occurs during the resistive linear phase of the process, in other words, that the transition time $t_R \sim \tau_A S^{1/7} \delta^{-4/7}$ [see Eq.(87)] exceeds the respective time $t_1 \sim \tau_A S^{1/3}$ in Eq.(39). This requirement puts an upper limit on the validity of the Rutherford regime, namely that $\delta \ll \delta_2 \sim S^{-1/3}$. Therefore, an issue of the strongly nonlinear forced reconnection, when the opposite inequality holds, i.e.

$$\delta \gg \delta_2 \sim S^{-1/3}, \qquad (90)$$

becomes of interest. It has been hypothesized (Wang & Bhattacharjee 1992) that in this case evolution of the system involves a phase of the Sweet-Parker type reconnection, which on the longer time-scale gives way to the Rutherford regime (see also Fitzpatrick 2003; Comisso, Grasso & Waelbroeck 2014, 2015). It turns out, however, that such a scenario actually does not take place. What happens instead is formation of the nonlinear CS which plays a central role in the subsequent evolution of the system (Vekstein & Kusano 2015). The point is that while in the case of the weak nonlinearity, when $\delta_1 \ll \delta \ll \delta_2$, the resistive effect comes into play before the nonlinear one, i.e. $t_1 < t_R$, in the case of (90) the nonlinearity comes first. Therefore, initial stages of the CS evolution can be analysed in the framework of the ideal MHD. Then, the first step is to find out when the nonlinearity interrupts the linear phase, which is described by Eqs. (36-38). According to these equations, inside the CS the flux- and stream-functions are equal to

$$\Psi_1 \sim \delta \cdot B_0 l \left(\frac{x}{l}\right)^2 \left(\frac{t}{\tau_A}\right), \qquad \Phi \sim \delta \cdot V_A l \left(\frac{x}{l}\right),$$

hence, the source term (83) is of order of

$$(\vec{\nabla}\Psi_1 \times \vec{\nabla}\Phi) \cdot \vec{z} \sim \delta^2 \cdot B_0 V_A \left(\frac{x}{l}\right)^2 \left(\frac{t}{\tau_A}\right) \qquad (91)$$

Furthermore, unlike the case of the resistive Rutherford regime discussed above, we are now in the realm of the ideal MHD, so the generating source (91) is balanced in Eq.(32a) by the temporal derivative of $\Psi_2^{(0)}$, while the resistive term is small (see below). Hence,

$$\frac{\partial \Psi_2^{(0)}}{\partial t} \sim \delta^2 \cdot B_0 V_A \left(\frac{x}{l}\right)^2 \left(\frac{t}{\tau_A}\right) \Rightarrow \Psi_2^{(0)} \sim \delta^2 \cdot B_0 l \left(\frac{x}{l}\right)^2 \left(\frac{t}{\tau_A}\right)^2 \qquad (92)$$

Therefore, the third-order torque (82) can be estimated as

$$T_3 \sim \frac{\Psi_1}{l} \cdot \frac{\Psi_2^{(0)}}{(\Delta x)^3} \sim \delta^3 \cdot \frac{B_0^2}{l^2} \left(\frac{t}{\tau_A}\right)^2,$$

which becomes comparable to the first-order torque (81) at $t \sim t_1^{(N)} \sim \tau_A \delta^{-1}$ [note that $t_1^{(N)} \ll t_1 \sim S^{1/3}$ under the condition (90)]. Regarding the ignored resistive term in Eq.(32a), it can be estimated as $\eta \partial^2 \Psi_2^{(0)} / \partial x^2 \sim \eta \delta^2 B_0 l^{-1} (t/\tau_A)^2$, while the source term (91) is of order of $\delta^2 B_0 V_A (\Delta x)^2 l^{-2} (t/\tau_A) \sim \delta^2 B_0 V_A (t/\tau_A)^{-1}$ [see Eq.(38)]. Hence, their ratio is of order of $(t/\tau_A)^3 \eta / V_A l \sim (t/\tau_A)^3 S^{-1}$, which is small indeed when $t \ll t_1$.

Thus, at $t \sim t_1^{(N)} \sim \tau_A \delta^{-1}$, when, according to (38), the CS thickness $\Delta x$ is of order of the boundary perturbation amplitude $a$, the system attains a state of equilibrium with the zero net torque. Therefore, in the absence of any plasma resistivity this would be the terminal ideal MHD equilibrium, where the nonlinear effect resolves the singularity present in the linear ideal MHD solution (18). If, however, a finite resistivity is present, this nonlinear CS has a finite life- time equal to the skin-time $(\Delta t)_S \sim (\Delta x)^2 / \eta \sim \tau_A S \delta^2$ (note that, according to (90), $(\Delta t)_S$ is much longer than $t_1^{(N)}$). At $t > t_1^{(N)}$ the reconnected flux, $\Psi_r(t) = \Psi(x=0, t)$ is increasing with time [see Eq. (32a)] as

$$\frac{d\Psi_r}{dt} \approx \eta \frac{\partial^2 \Psi}{\partial x^2} \sim \eta \frac{B_0 \delta}{\Delta x} \sim \eta \frac{B_0}{l} \Rightarrow \Psi_r(t) \sim B_0 l \frac{t}{\tau_\eta} \sim B_0 l S^{-1} \frac{t}{\tau_A}$$

Therefore, the width of the respective magnetic islands is equal to

$$w_r(t) \sim [l \Psi_r(t) / B_0]^{1/2} \sim l S^{-1/2} (t/\tau_A)^{1/2} \qquad (93)$$

As seen from (93), $w_r$ remains small compared to $\Delta x \sim l\delta$ when $t < (\Delta t)_S$, and the two become equal at $t \sim (\Delta t)_S$. It means that the subsequent forced reconnection takes place in the considered above constant-$\psi$ Rutherford regime with the reconnection time-scale equal to $\tau_R \sim \tau_A S \delta^{1/2}$ [see Eq.(87)].

However, realization of such a "conventional" scenario, which is associated with the three consecutive time-scales $t_1^{(N)} << (\Delta t)_S << \tau_R$, is unlikely because of the secondary tearing (plasmoid) instability (Biskamp 1986; Loureiro, Schekochihin & Cowley 2007) of the formed nonlinear CS. Involvement of this instability in the process of forced magnetic reconnection was first observed in the numerical simulation (Comisso, Grasso & Waelbroeck 2014). This followed with a complete analytical theory of the nonlinear forced reconnection and onset of plasmoid instability (Vekstein & Kusano 2015). The point is that the nonlinear CR under discussion is tearing unstable, and this instability is fast enough to develop before the transition to the Rutherford phase occurs.

In order to demonstrate this, it is useful to start with a brief summary of the resistive tearing instability of some general CS of length $L$, thickness $h << L$, and magnetic field $B$. These parameters define the respective Alfven velocity $V_A^{(h)}$, the Alfven transit time $\tau_A^{(h)} = h/V_A^{(h)}$, and the Lundquist number $S_h = hV_A^{(h)}/\eta$, which is assumed to be large: $S_h >> 1$. Such a CS is usually tearing unstable with respect to modes with a wave-length (along the CS) $\lambda > h$. There are two regimes of this instability. If the wave-length is not too long, namely $h < \lambda < \lambda_* \sim h S_h^{1/4}$, it corresponds to the constant-$\psi$ regime (Furth, Killeen & Rosenbluth 1963), and the respective instability growth-rate scales as

$$\gamma_h \sim (\tau_A^{(h)})^{-1} S_h^{-3/5} (\lambda/h)^{2/5}, \qquad (94)$$

hence, increasing with $\lambda$. On the other hand, in the non-constant-$\psi$ case, which takes place when $\lambda > \lambda_*$, the growth rate goes down for longer wave-lengths as $\gamma_h \sim (\tau_A^{(h)})^{-1} S_h^{-1/3} (\lambda/h)^{-2/3}$ (Coppi et al 1976). Therefore, the most unstable mode (a one with the maximum growth rate) has a wave-length $\lambda \sim \lambda_*$, and the growth-rate $\gamma_h^{(\max)} \sim (\tau_A^{(h)})^{-1} S_h^{-1/2}$.

Now these results can be applied to the nonlinear CS under discussion, for which $L = l, h = a = l\delta$, and $B = B_0(a/l) = B_0\delta$. Under these parameters one gets $V_A^{(h)} = \tilde{V}_A \sim V_A \delta$, $\tilde{\tau}_A \sim \tau_A$, and $\tilde{S} \sim S\delta^2$ (note that, due to inequality (90), $\tilde{S} \gg 1$). Then, $\tilde{\lambda}_* \sim a\tilde{S}^{1/4} \sim lS^{1/4}\delta^{3/2}$, and this most unstable mode fits in the CS of length $L \sim l$, i.e. $\tilde{\lambda}_* < l$, if $\delta < S^{-1/6}$. In this case the respective growth-rate is $\gamma_{\max} \sim \tilde{\tau}_A^{-1}\tilde{S}^{-1/2} \sim \tau_A^{-1}S^{-1/2}\delta^{-1}$, which is sufficient for the instability development during the CS life-time $(\Delta t)_S \sim \tau_A S\delta^2$ (indeed, the product $\gamma_{\max} \cdot (\Delta t)_S \sim S^{1/2}\delta \gg 1$). Furthermore, it also allows to predict the number of plasmoids, $N_{pl}$, expected to form during the initial linear stage of the plasmoid instability: $N_{pl} \sim L/\tilde{\lambda}_* \sim S^{-1/4}\delta^{-3/2} \gg 1$. If, however, $\delta > S^{-1/6}$, the fastest mode fitting in the CS is the one with $\lambda \sim l < \tilde{\lambda}_*$. Its growth-rate, according to (94), is equal to $\gamma \sim \tilde{\tau}_A^{-1}\tilde{S}^{-3/5}(l/a)^{2/5} \sim \tau_A^{-1}S^{-3/5}\delta^{-8/5}$. It is still fast enough, as $\gamma \cdot (\Delta t)_S \sim (S\delta)^{2/5} \gg 1$, but in this case the number of the initially generated plasmoids is just a few: $N_{pl} \sim 1$. Such analytical predictions are in a very good agreement with the numerical results reported in Comisso et al (2014). Note, however, that since this numerical simulation corresponds to the case of the visco-resistive forced reconnection, its direct comparison with the analytical theory requires the respective extension of the theory (see Vekstein & Kusano 2015).

In the Hall-MHD framework the situation is even more favourable to the plasmoid instability development (Vekstein & Kusano 2017). Indeed, the Hall effect makes the secondary tearing instability faster by providing additional inflow of magnetic flux into the reconnection site, but it leaves intact the nonlinear CS resistive life-time $(\Delta t)_S$. Moreover, in the Hall-MHD case the onset of plasmoid instability may occur also during the resistive phase of the CS evolution (Vekstein & Kusano 2017). This is not possible in the standard single-fluid MHD because the system slips into the Rutherford regime due to halting of the plasma flow. On the contrary, in the respective Hall-mediated phase this effect becomes irrelevant because advection of the poloidal magnetic field is provided by the Hall electric current rather than by the bulk flow of the plasma.

The issue of the secondary tearing instability is presently a hot topic in the magnetic reconnection research because nonlinear dynamics of generated plasmoids paves the way to fast magnetic reconnection (Brattacharjee et al 2009; Uzdensky et al 2010; Shibayama et al 2015). It attracted a significant number of publications, most of which are numerical simulations. The problem is that a self-consistent description of this process is not a simple task: it requires following an entire CS evolution which, at some point, brings about the onset of the plasmoid instability (Uzdensky & Loureiro 2016). However, in order to get deeper understanding of the issue, in particular, its scaling with the plasma and the magnetic field parameters, one needs some tractable analytical models. Therefore, in this respect the discussed Taylor model of forced reconnection provides a very useful tool.

**Acknowledgements**

The author is grateful to Mykola Gordovskyy for help in preparing this article.